\documentclass{article}

\usepackage{arxiv}

\usepackage[utf8]{inputenc} 
\usepackage[numbers]{natbib}
\usepackage{hyperref}       
\usepackage{url}            
\usepackage{booktabs}       
\usepackage{amsfonts}       
\usepackage{nicefrac}       
\usepackage{microtype}      
\usepackage{lipsum}
\usepackage{graphicx}
\usepackage{multirow}
\usepackage{amsmath}
\usepackage{subcaption}

\title{Impact of lung segmentation on the diagnosis and explanation of COVID-19 in chest X-ray images}

\author{
 Lucas Teixeira\\
  Universidade Estadual de Maringá\\
   \And
 Rodolfo M. Pereira \\
  Instituto Federal do Paraná\\
  \And
 Diego Bertolini \\
  Universidade Tecnológica Federal do Paraná\\
  \And
 Luiz S. Oliveira \\
  Universidade Federal do Paraná\\
  \And
 Loris Nanni \\
  Università Degli Studi di Padova\\
  \And
 George D. C. Cavalcanti \\
  Universidade Federal de Pernambuco\\
  \And
 Yandre M. G. Costa \\
  Universidade Estadual de Maringá\\
}

\begin{document}
\maketitle
\begin{abstract}
COVID-19 frequently provokes pneumonia, which can be diagnosed using imaging exams. Chest X-ray (CXR) is often useful because it is cheap, fast, widespread, and uses less radiation. Here, we demonstrate the impact of lung segmentation in COVID-19 identification using CXR images and evaluate which contents of the image influenced the most. Semantic segmentation was performed using a U-Net CNN architecture, and the classification using three CNN architectures (VGG, ResNet, and Inception). Explainable Artificial Intelligence techniques were employed to estimate the impact of segmentation. A three-classes database was composed: lung opacity (pneumonia), COVID-19, and normal. We assessed the impact of creating a CXR image database from different sources, and the COVID-19 generalization from one source to another. The segmentation achieved a Jaccard distance of 0.034 and a Dice coefficient of 0.982. The classification using segmented images achieved an F1-Score of 0.88 for the multi-class setup, and 0.83 for COVID-19 identification. In the cross-dataset scenario, we obtained an F1-Score of 0.74 and an area under the ROC curve of 0.9 for COVID-19 identification using segmented images. Experiments support the conclusion that even after segmentation, there is a strong bias introduced by underlying factors from different sources.
\end{abstract}


\section{Introduction}
\label{intro}

The Coronavirus disease 2019 (COVID-19) pandemic, caused by the virus named Severe Acute Respiratory Syndrome Coronavirus 2 (SARS-CoV-2), has become the most significant public health crisis our society has faced recently \footnote{https://covid19.who.int/}. COVID-19 affects mainly the respiratory system and, in extreme cases, causes a massive inflammatory response that reduces the total lung capacity \citep*{tay2020trinity}. COVID-19 high transmissibility, lack of general population immunization, and high incubation period \citep*{lauer2020incubation} makes it a dangerous and lethal disease. In these circumstances, artificial intelligence (AI) based solutions are being used in various contexts, from diagnostic support to vaccine development \citep*{alimadadi2020artificial}.

The standard imaging tests for pneumonia, and consequently COVID-19, are chest X-ray (CXR) and computed tomography or computerized X-ray imaging (CT) scan. The CT scan is the gold standard for lung disease diagnosis since it generates very detailed images. However, CXR is still very useful in particular scenarios, since they are cheaper, generate the resulting images faster, expose the patient to much less radiation, and it is more widespread in the emergency care units \citep*{self2013high}.

After the COVID-19 outbreak, several studies were proposed to investigate its diagnostic based on the use of images taken from the lungs \citep*{pereira2020covid, wang2020covid}. Despite the impressive advances, there is a lack of more critical analysis regarding the content captured in those images that contribute to consistent results \citep*{maguolo2021critic, cruz2020composition, tartaglione2020unveiling}. The results reported by \citep*{maguolo2021critic} were one of the main reasons we decided to evaluate the impact of lung segmentation in COVID-19 identification. A proper lung segmentation might mitigate the bias introduced by composing multiple databases and provides a more realistic performance.

Our main objective is to evaluate the impact of lung segmentation in identifying pneumonia caused by different microorganisms using CXR images obtained from various sources (i.e., Cohen, RSNA pneumonia detection challenge, among others). We have primarily focused on CXR images due to their smaller cost and high availability in the emergency care units, especially those located in less economically developed regions. Moreover, we emphasize COVID-19, aiming to provide solutions that can be useful in the current pandemic context. To support that objective, we used an U-Net Convolutional Neural Network (CNN) for lung segmentation, and three popular CNN models for COVID-19 identification: VGG16 \citep*{simonyan2014very}, ResNet50V2 \citep*{he2016identity} and InceptionV3 \citep*{szegedy2016rethinking}. Since our main goal is to highlight the importance of lung segmentation and not claim state-of-art COVID-19 identification, we preferred to use popular, consolidated, and well-established CNN architectures. Furthermore, to provide a more complete and realistic overview, we also evaluated specific scenarios to assess the database bias, i.e., the importance of the image source for the classification model and COVID-19 generalization, i.e., the usage of COVID-19 images from one database to train a classification model to identify COVID-19 cases in a different database, which represents the less biased scenario evaluated in this paper.

We first improved our previously created COVID-19 database (i.e., RYDLS-20 \citep*{pereira2020covid}), now called RYDLS-20-v2, adding more image sources. Then, we set up the problem as a multi-class classification problem with three classes: lung opacity, COVID-19, and normal lungs (i.e., no-pneumonia), in which lung opacity means pneumonia caused by any previously known pathogen. We decided to use three classes because there is a considerable difference between COVID-19 and healthy patients, and a binary classification problem might not be challenging enough; hence we added a confounding class containing pneumonia caused by any other pathogen, except COVID-19. To segment lung images, we applied a deep learning approach using a U-Net CNN architecture \citep*{ronneberger2015u}.

Over the last few years, the area known as Explainable Artificial Intelligence (XAI) has attracted many researchers in the artificial intelligence (AI) field. The main interest of XAI is to research and develop approaches to explain the individual predictions of modern machine learning (ML) based solutions. In medical applications based on images, we understand that a proper explanation regarding the obtained decision is fundamental. In an ideal scenario, the decision support system should be able to suggest the diagnosis and justify, as better as possible, which contents of the image have decisively contributed to achieving a particular decision.

To assess the impact of lung segmentation on the identification of COVID-19, we used two XAI approaches: Local Interpretable Model-agnostic Explanations (LIME) \citep*{ribeiro2016should} and Gradient-weighted Class Activation Mapping (Grad-CAM) \citep*{selvaraju2017grad}. LIME works by finding features, superpixels (i.e., particular zones of the image), that increases the probability of the predicted class, i.e., regions that support the current model prediction. Such regions can be seen as important regions because the model actively uses them to make predictions. Grad-CAM focuses on the gradients flowing into the last convolutional layer of a given CNN for a specific input image and label. We can then visually inspect the activation mapping (AM) to verify if the model is focusing on the appropriate portion of the input image. Both techniques are somewhat complementary, and by exploring them, we can provide a more complete report of the lung segmentation impact on COVID-19 identification.

Our results indicated that when the whole image is considered, the model may learn to use other features besides lung opacities, or even from outside the lungs region. In such cases, the model is not learning to identify pneumonia or COVID-19, but something else. Thus, we can infer that the model is not reliable even though it achieves a good classification performance. Using lung segmentation, we would supposedly remove a meaningful part of noise and background information, forcing the model to take into account only data from the lung area, i.e., desired information in this specific context. Thus, the classification performance in models using segmented CXR images tends to be more realistic, closer to human performance, and better reasoned.

The remaining of this paper is organized as follows: Section \ref{sec:method} introduces our proposed methodology and experimental setup. Section \ref{sec:related} presents current studies about COVID-19 identification and discusses about the state-of-art. Section \ref{sec:method} shows details about our methodology, experimental setup, including database, algorithms, and parameters. Section \ref{sec:results} presents the obtained results. Later, Section \ref{sec:discussion} discusses the obtained results. Finally, Section \ref{sec:conclusion} presents our conclusions and possibilities for future works.

\section{Related Works}
\label{sec:related}

This section discusses some influential papers in the literature related to one of the following topics: model inspection and explainability in lung segmentation or COVID-19 identification in CXR/CT images. Moreover, we also discuss potential limitations, biases, and problems of COVID-19 identification given the current state of available databases.

It is important to observe that as the identification of COVID-19 in CXR/CT images is a hot topic nowadays due to the growing pandemic, it is unfeasible to represent the actual state-of-the-art for this task since new works are emerging every day. Nevertheless, we may observe that most of those works aim to investigate configurations for Deep Neural Networks, which is already different from our proposal.

In order to show how fast is growing the research content around the topic of Machine Learning applications on COVID-19, we can briefly present some surveys and reviews published in the literature. Still, in April 2020, \citep*{shi2020review} already presented one of the firsts reviews of techniques to perform COVID-19 detection in X-ray and CT-Scan images, aiming at tasks such as screening process and severity assessment. Recently, \citep*{bhattacharya2021deep} and \citep*{islam2021review} presented surveys focused on challenges, issues and future research directions related to deep learning implementations for COVID-19 detection. Moreover, \citep*{roberts2021common} and \citep*{santa2021public} presented critical systematic reviews of COVID-19 automatic detection focused on the potential clinical use of the proposed techniques.

In this field of investigation, the works are typically accomplished using deep learning models. Deep learning models usually tend to produce results that cannot be naturally explained by themselves. It happens due to the high complexity of these models. Aiming to overcome this issue and trying to open the ``black-box'' characterized by these models, XAI techniques have been more used to search for more convincing shreds of evidence that could help to understand why an AI system gave a particular response. By analyzing the literature, we noticed some works somehow related to this one because they evaluated deep models using lung images for COVID-19 detection in an XAI perspective.

In this sense, \citep*{DBLP:journals/corr/abs-2104-14506} used CAM, LIME, and SHAP as XAI techniques to provide more granular information to support clinician's decision making in the context of COVID-19 classification starting from chest CT scanned images. For this purpose, the authors trained the models using private databases composed of images taken from four Chinese hospitals and tested them on the open-access CC-CCII dataset \citep*{ZHANG20201423}, a publicly available dataset. The authors concluded that the XAI enhanced classifier was able to provide robust classification results and also a convincing explanation about them.

\citep*{BRUNESE2020105608} proposed a method composed of three steps aiming to detect lung diseases and to provide a kind of explanation regarding the decision obtained. Experiments were conducted on two datasets with a total of 6523 CXR images. The steps which compose the proposal can be summarized as follows: i) in the first step, the method performs the discrimination between a healthy and a chest X-ray related to pulmonary diseases in general; ii) in the second step, the method performs the discrimination between COVID-19 pneumonia and pneumonia provoked by other diseases; iii) in the third and last step, the method tries to present some explanation about the decision taken. For this, samples of chest X-rays highlighting the fundamental regions in the X-ray for COVID-19 prediction are provided.

From this point, we focus on works devoted to COVID-19 identification using chest images that somehow dealt with the identification of regions of interest. \citep*{Wang_2021} proposed a joint deep learning model of 3D lesion segmentation and classification for diagnosing COVID-19. For this purpose, they created a large-scale CT database containing 1,805 3D CT scans with fine-grained lesion annotations. The authors' main idea was to explore the inherent correlation between the 3D lesion segmentation and disease classification. The authors concluded that the joint learning framework proposed could significantly improve both the performance of 3D segmentation and disease classification in terms of efficiency and efficacy.

\citep*{wang2021deep} created a deep learning pipeline for the diagnosis and discrimination of viral, non-viral, and COVID-19 pneumonia, composed of a CXR standardization module followed by a thoracic disease detection module. The first module (i.e., standardization) was based on anatomical landmark detection. The landmark detection module was trained using 676 CXR images with 12 anatomical landmarks labeled. Three different deep learning models were implemented and compared (i.e., U-Net, fully convolutional networks, and DeepLabv3). The system was evaluated in an independent set of 440 CXR images, and the performance was comparable to senior radiologists.

In \citep*{chen2020residual}, the authors proposed an automatic segmentation approach using deep learning (i.e., U-Net) for multiple regions of COVID-19 infection. In this work, a public CT image dataset was used with 110 axial CT images collected from 60 patients. The authors describe the use of Aggregated Residual Transformations and a soft attention mechanism in order to improve the feature representation and increase the robustness of the model by distinguishing a greater variety of symptoms from the COVID-19. Finally, an excellent performance on COVID-19 chest CT image segmentation was reported in the experimental results.

In \citep*{DeGrave} the authors investigate if the high rates presented in COVID-19 detection systems from chest radiographs using deep learning may be due to some bias related to shortcut learning. Using explainable artificial intelligence (AI) techniques and generative adversarial networks (GANs), it was possible to observe that systems that presented high performance end up employing undesired shortcuts in many cases. The authors evaluate techniques in order to alleviate the problem of shortcut learning. \citep*{DeGrave} demonstrates the importance of using explainable AI in clinical deployment of machine-learning healthcare models to generate more robust and valuable models.

\citep*{bassi2021covid19} present segmentation and classification methods using deep neural networks (DNNs) to classify chest X-rays as COVID-19, normal, or pneumonia. U-Net architecture was used for the segmentation and DenseNet201 for classification. The authors employ a small database with samples from different locations. The main goal is to evaluate the generalization of the generated models. Using Layer-wise Relevance Propagation (LRP) and the Brixia score, it was possible to observe that the heat maps generated by LRP show that areas indicated by radiologists as potentially important for symptoms of COVID-19 were also relevant for the stacked DNN classification. Finally, the authors observed that there is a database bias, as experiments demonstrated differences between internal and external validation.

Following this context, after \citep*{cohen2020covid} started putting together a repository containing COVID-19 CXR and CT images, many researchers started experimenting with automatic identification of COVID-19 using only chest images. Many of them developed protocols that included the combination of multiple chest X-rays database and achieved very high classification rates, much higher than human performance \citep*{kanne2020essentials}. Moreover, there have been multiple reports in the literature that supports the fact that many published papers might have used biased testing protocols, which resulted in unrealistic results \citep*{maguolo2021critic, cruz2020composition, tartaglione2020unveiling, cohen2020limits, lopez2021current}.

Although the literature on the subject addressed here is very recent, we notice an increasing concernment regarding the explainability of the results obtained, thanks to the seriousness and urgency of this matter. Even though there are other works exploring XAI on COVID-19 detection using CXR images, as far as we know, at the time of this publication none of them explored exactly the same protocol we explore here, considering both the segmentation of the regions of interest followed by classification supported by XAI.

\section{Material and Methods}
\label{sec:method}

We focused on exploring data from CXR images for reliable identification of COVID-19 among pneumonia caused by other micro-organisms. Hence, we proposed a specific method that allowed us to assess lung segmentation's impact on COVID-19 identification.

\begin{figure}[htbp!]
\centerline{\includegraphics[width=\columnwidth]{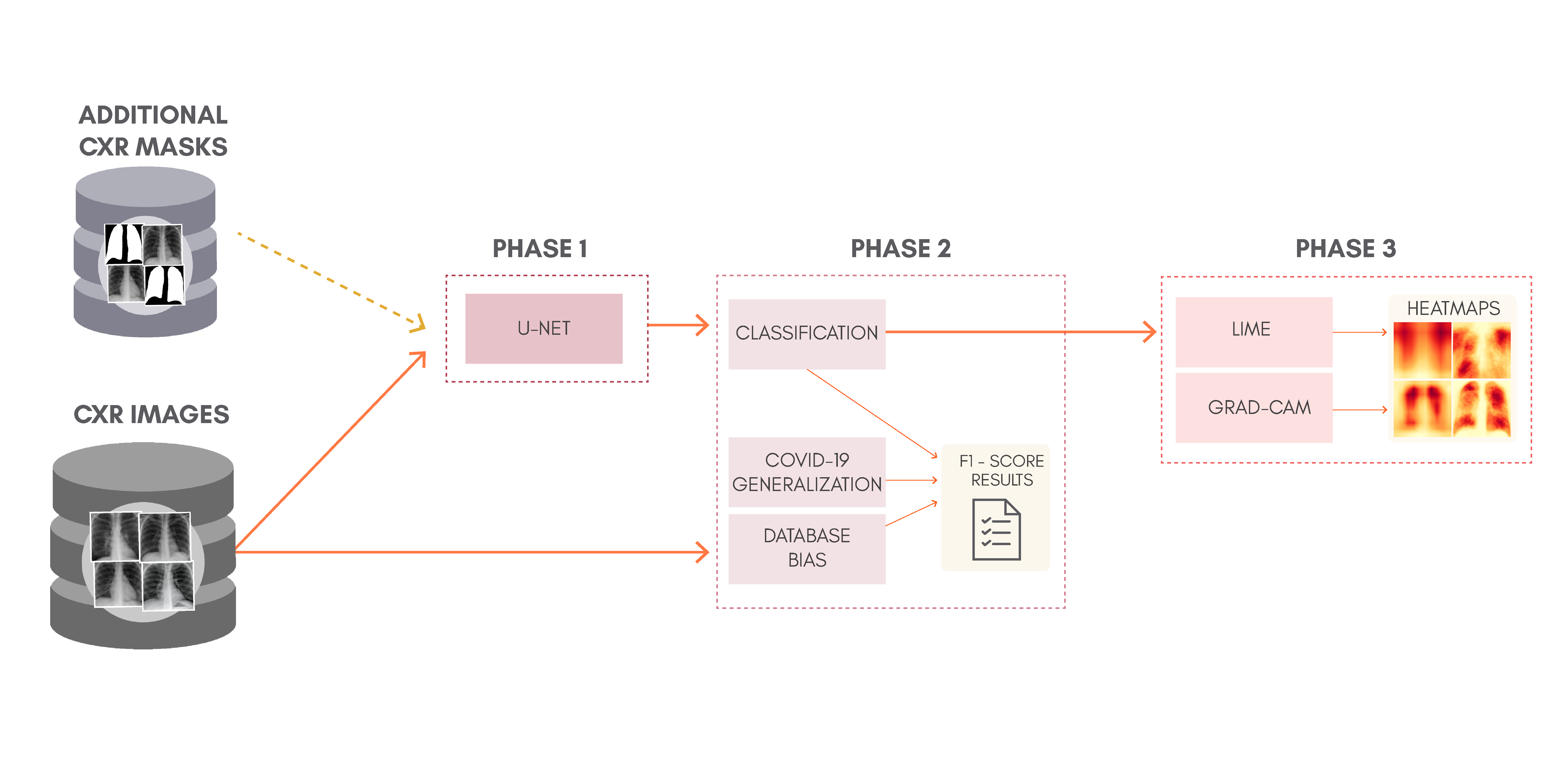}}
\caption{Proposed methodology.}
\label{fig:method}
\end{figure}

To better understand the proposal of this work, Figure \ref{fig:method} shows a general overview of the classification approach adopted, containing: lung segmentation (Phase 1), classification (Phase 2), and XAI (Phase 3). Phase 1 is skipped entirely for the classification of non-segmented CXR images. Although simple, this can be considered as a kind of ablation study since we isolate the lung segmentation phase and evaluate its impact. In order to allow the reproduction of our exact experiments, we made all our code and database available in a GitHub repository\footnote{https://github.com/lucasxteixeira/covid19-segmentation-paper}.

\subsection{Lung Segmentation (Phase 1)}

\begin{figure}[htbp!]
\centering
\begin{subfigure}{0.32\textwidth}
  \centering
  \includegraphics[width=.98\textwidth]{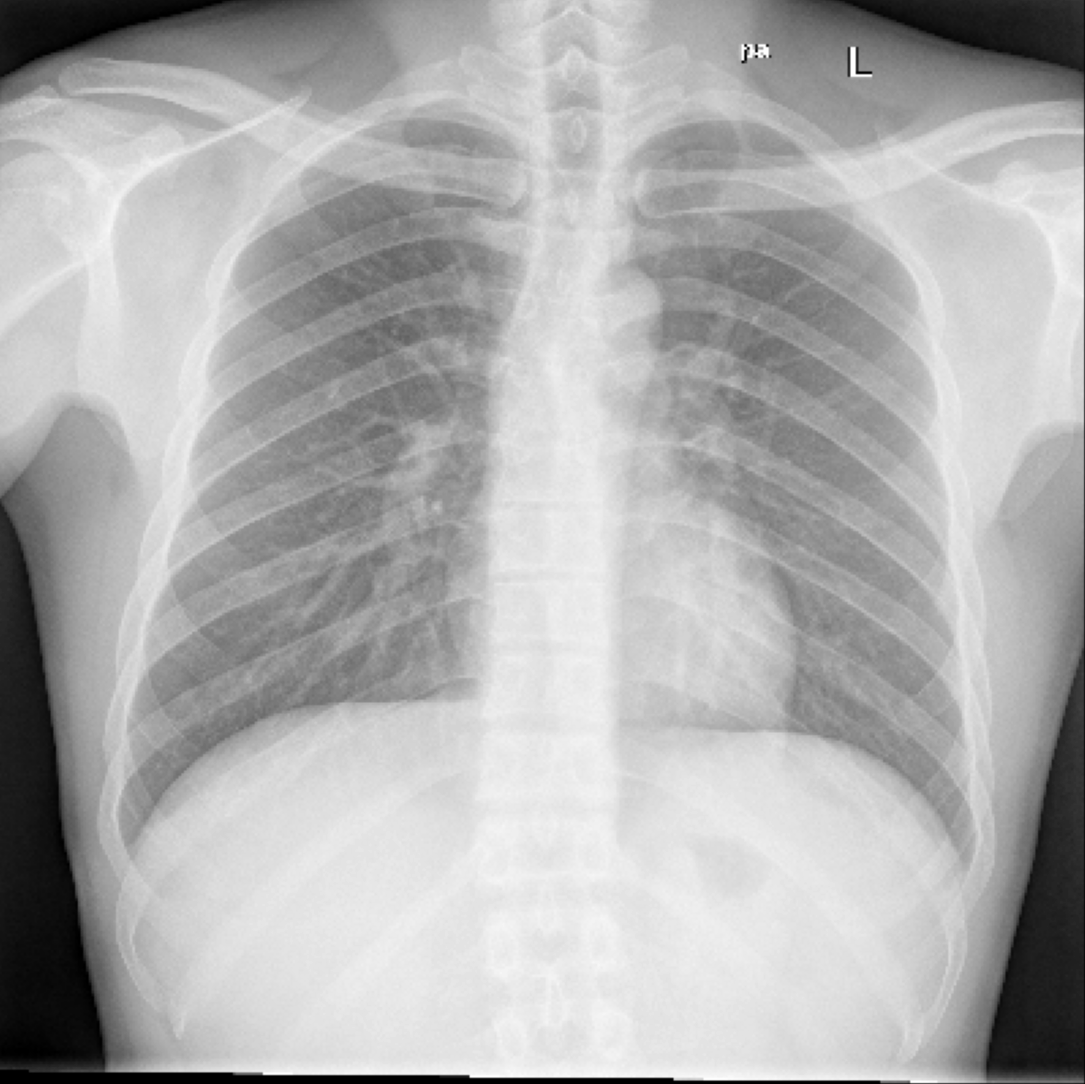}
  \caption{CXR image.}
\end{subfigure}
\begin{subfigure}{0.32\textwidth}
  \centering
  \includegraphics[width=.98\textwidth]{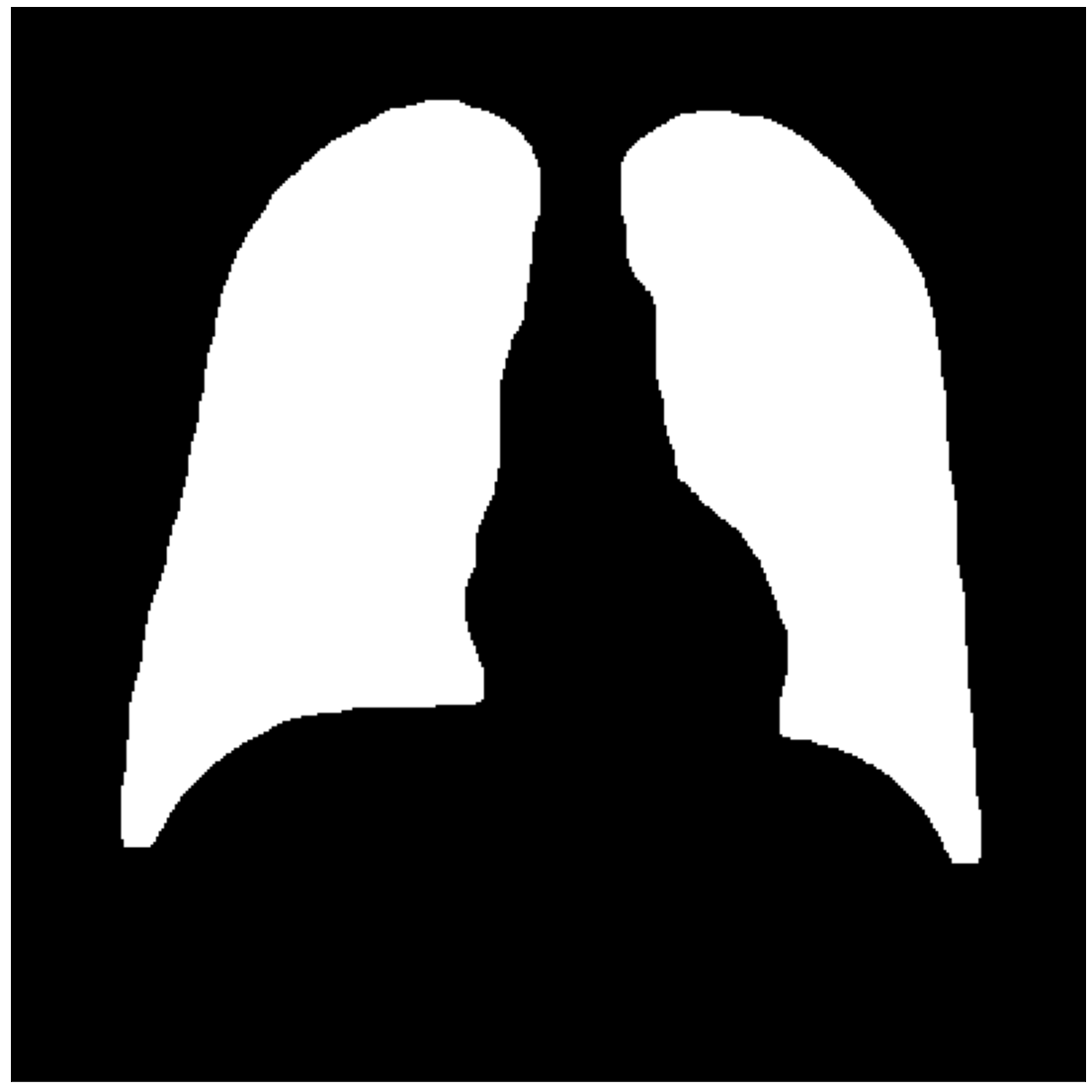}
  \caption{Binary mask.}
\end{subfigure}
\begin{subfigure}{0.32\textwidth}
  \centering
  \includegraphics[width=.98\textwidth]{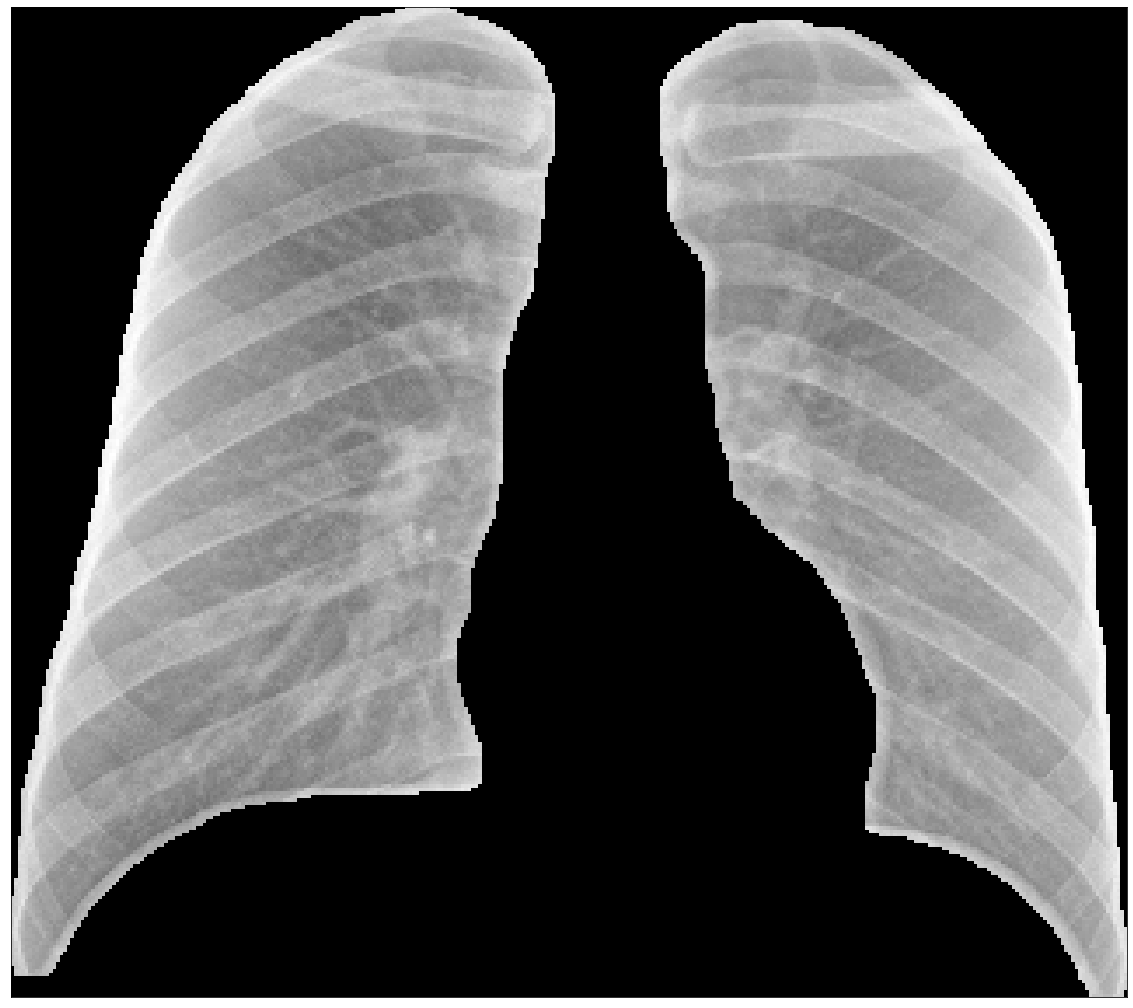}
  \caption{Segmented lungs.}
\end{subfigure}
\caption{Lungs segmentation on CXR image.}
\label{fig:segmentation}
\end{figure}

The first phase in our method is the lung segmentation, aiming to remove all background and retain only the lung area. We expect it to reduce noise that can interfere with the model prediction. Figure \ref{fig:segmentation} presents an example of lung segmentation.

\begin{figure}[htbp!]
\centering
\begin{subfigure}{0.4\textwidth}
  \centering
  \includegraphics[width=.98\textwidth]{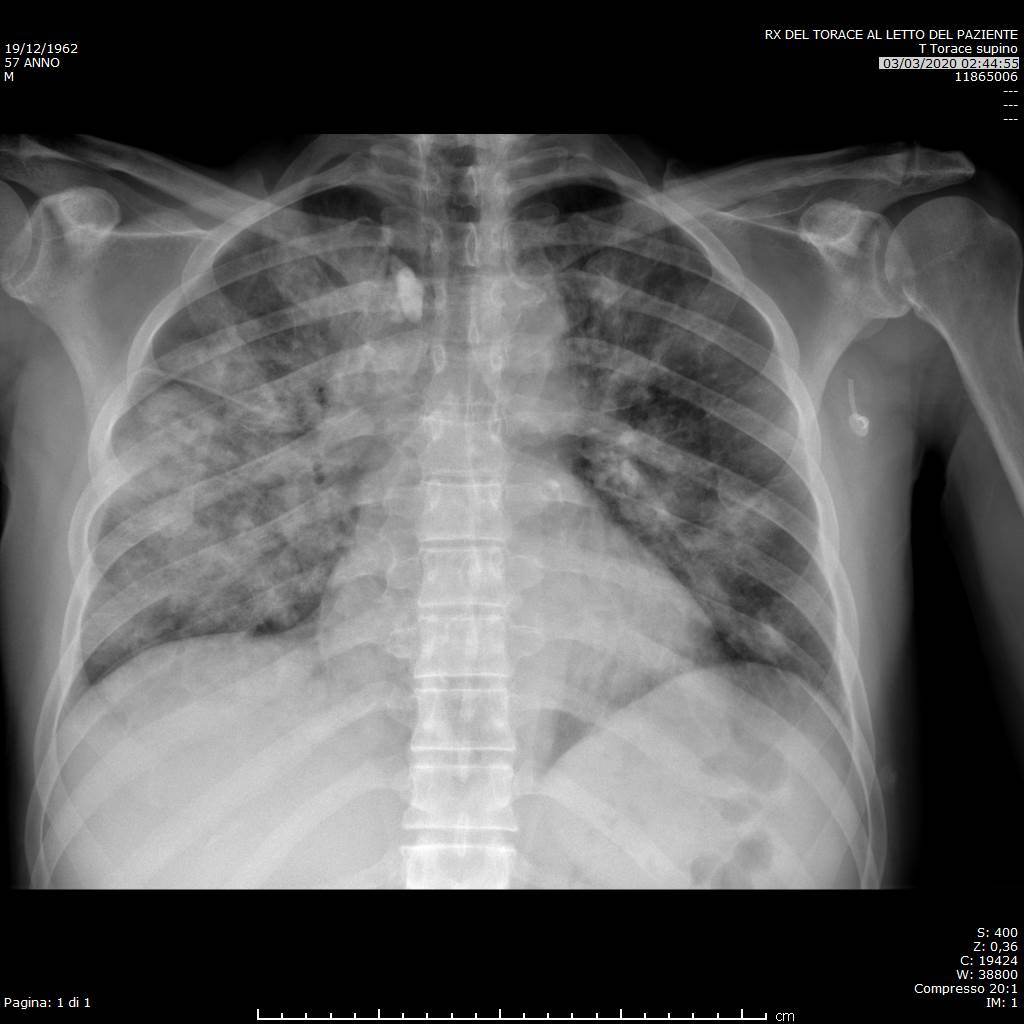}
\end{subfigure}
\begin{subfigure}{0.4\textwidth}
  \centering
  \includegraphics[width=.98\textwidth]{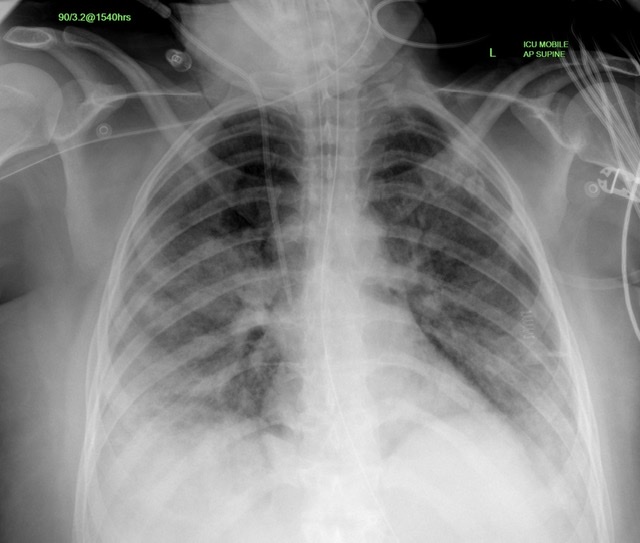}
\end{subfigure}
\caption{CXR with burned in annotations.}
\label{fig:cxr-burnedin}
\end{figure}

Specifically, in deep models, any extra information can lead to model overfitting. This is especially important in CXR since many images contain burned-in annotations about the machine, operator, hospital, or patient. Figure \ref{fig:cxr-burnedin} presents an example of CXR images with burned in information.

We expect that the models using segmented images rely on information in the lung area rather than background information, i.e., an increase in the model reliability and prediction quality in a real-world scenario. For example, if a model is trained to predict lung opacity, it must use lung area information. Otherwise, it is not identifying opacity but something else.

In order to perform lung segmentation, we applied a CNN approach using the U-Net architecture \citep*{ronneberger2015u}. The U-Net input is the CXR image, and the output is a binary mask that indicates the region of interest (ROI). Thus, the training requires a previously set of binary masks.

The COVID-19 dataset used does not have manually created binary masks for all images. Thus, we adopted a semi-automated approach to creating binary masks for all CXR images. First, we used three additional CXR datasets with binary masks to increase the training sample size and some binary masks provided by v7labs\footnote{https://github.com/v7labs/COVID-19-xray-dataset}. We then trained the U-Net model and used it to predict the binary masks for all images in our dataset. After that, we reviewed all predicted binary masks and manually created masks for those CXR images that the model was unable to generalize well. We repeated this process until we judged the result satisfactory and achieved a good intersection between target and obtained regions.

\subsubsection{Lung Segmentation Database}

Table \ref{table:db-segmentation} presents the main characteristics of the database used to perform experimentation on lung segmentation. It comprises 1,645 CXR images, with a 95/5 percentage train/test split. In addition, we also created a third set for training evaluation, called validation set, containing 5 percent of the training data. Lung segmentation is trying to predict a binary mask indicating the lung region, irrespective of the input class (COVID-19, lung opacity, or healthy patients). Therefore, the class distribution has little impact on the outcome. Thus, we decided to use a random holdout split for validation.

{
\begin{table}[htbp!]
\centering
\small
\caption{Lung segmentation database.}
\begin{tabular}{@{}lll@{}}
\toprule
\textit{Characteristic} & \textit{Samples} \\ \midrule
Train & 1,483 \\
Validation & 79 \\
Test & 83 \\ \midrule
Total & 1,645 \\ \bottomrule
\end{tabular}
\label{table:db-segmentation}
\end{table}
}

Table \ref{table:db-segmentation-distribution} presents the samples distribution for each source.

{
\begin{table}[htbp!]
\centering
\small
\caption{Lung segmentation database composition.}
\begin{tabular}{@{}lll@{}}
\toprule
\textit{Source} & \textit{Samples} \\ \midrule
Cohen v7labs\footnote{https://github.com/v7labs/covid-19-xray-dataset} & 489 \\
Montgomery & 138 \\
Shenzhen & 566 \\
JSRT & 247 \\
Manually created & 205 \\ \bottomrule
\end{tabular}
\label{table:db-segmentation-distribution}
\end{table}
}

\subsubsection{U-Net}

The U-Net CNN architecture is a fully convolutional network (FCN) that has two main components: a contraction path, also called an encoder, which captures the image information; and the expansion path, also called decoder, which uses the encoded information to create the segmentation output \citep*{ronneberger2015u}.

\begin{figure}[htbp!]
\centerline{\includegraphics[width=0.9\columnwidth]{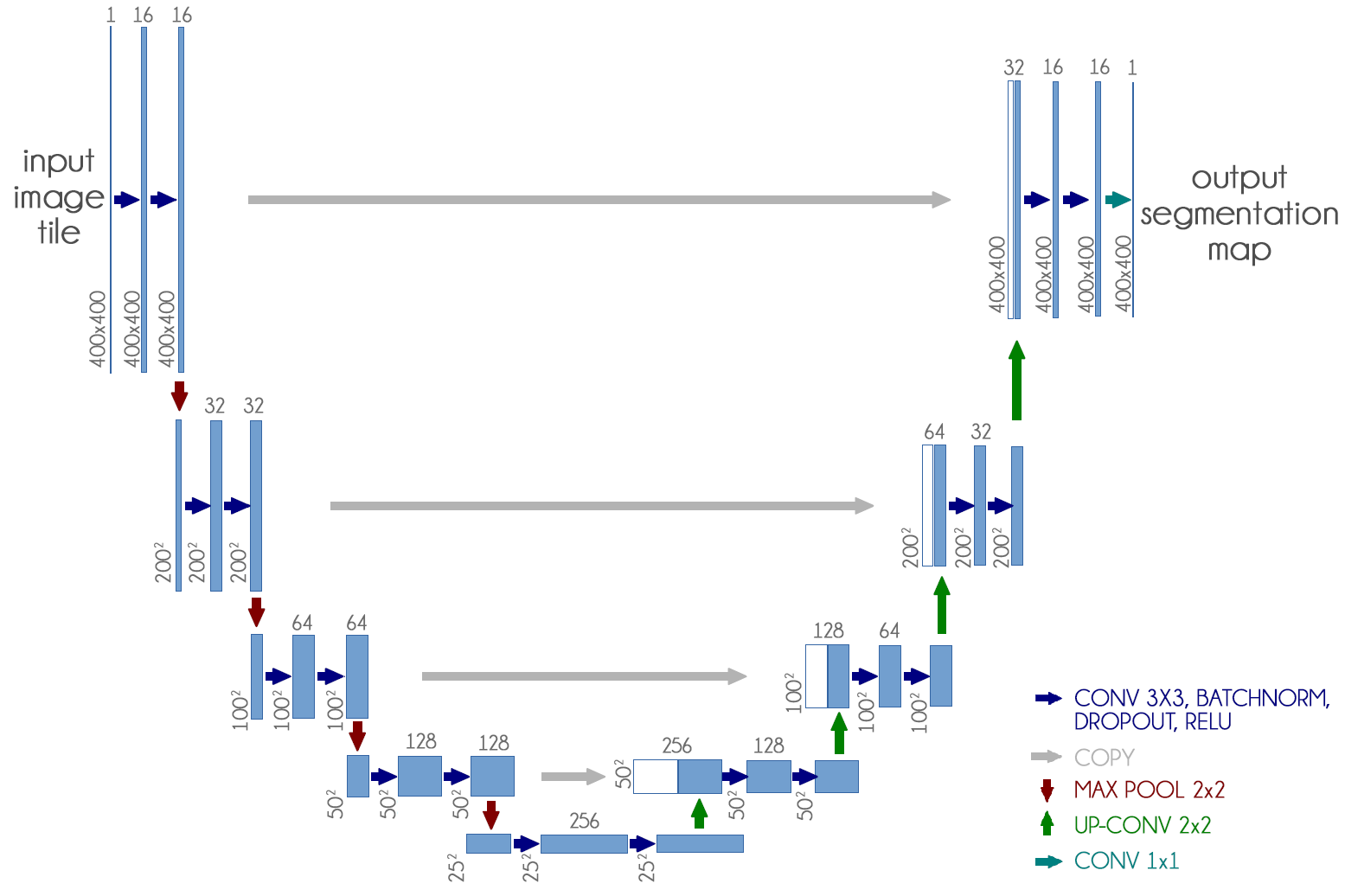}}
\caption{Custom U-Net architecture}
\label{fig:unet-custom}
\end{figure}

We used the U-Net CNN architecture with some small changes: we included dropout and batch normalization layers in each contracting and expanding block. These additions aim to improve training time and reduce overfitting. Figure \ref{fig:unet-custom} presents our adapted U-Net architecture.

Furthermore, since our dataset is not standardized, the first step was to resize all images to 400px $\times$ 400px, because it presented a good balance between computational requirements and classification performance. We also experimented with smaller and larger dimensions with no significant improvement.

In this model, we achieve a much better result without using transfer learning and training the network weights from scratch. Table \ref{table:unet-parameter} reports the parameters used in U-Net training.

{
\begin{table}[htbp!]
\centering
\small
\caption{U-Net parameters.}
\begin{tabular}{@{}ll@{}}
\toprule
\textit{Parameter} & \textit{Value} \\ \midrule
Epochs & 100 \\
Batch size & 16 \\
Learning rate & 0.001 \\ \bottomrule
\end{tabular}
\label{table:unet-parameter}
\end{table}
}

After the segmentation, we applied a morphological opening with 5 pixels to remove small brights spots, which usually happened outside the lung region. We also applied a morphological dilation with 5 pixels to increase and smooth the predicted mask boundary. Finally, we also cropped all images to keep only the ROI indicated by the mask. After crop the images were also resized to 300px $\times$ 300px. Figure \ref{fig:segmentation} shows an example of this process.

Besides, we also applied data augmentation techniques extensively to further expand our training data. Details regarding the usage and parameters will be discussed in Section \ref{sec:aug}.

\subsection{Classification (Phase 2)}

We chose a simple and straightforward approach with three of the most popular CNN architectures: VGG16, ResNet50V2 InceptionV3. For all, we applied transfer learning by loading pre-trained weights from ImageNet only for the convolutional layers \citep*{gulli2017deep}. We then added three fully-connected (FC) layers together, followed by dropout and batch normalization layers containing 1024, 1024, and 512 units. We performed the classification using full and segmented CXR images independently.

Furthermore, we also evaluated two specific scenarios to assess any bias in our proposed classification schema. First, we built a specific validation approach to assess the COVID-19 generalization from different sources, i.e., we want to answer the following question: is it possible to use COVID-19 CXR images from one database to identify COVID-19 in another different database? This scenario is one of our main contributions since it represent the least database biased scenario.

Then, we also evaluated a database classification scenario, in which we used the database source as the final label, and used full and segmented CXR images to verify if lung segmentation reduces the database bias. We want to answer the following question: does lung segmentation reduces the underlying differences from different databases which might bias a COVID-19 classification model?

In the literature, many papers employ complex classification approaches. However, a complex model does not necessarily mean better performance whatsoever. Even very simple deep architectures tend to overfit very quickly \citep*{srivastava2014dropout}. There must be a solid argument to justify applying a complicated approach to a low sample size problem. Additionally, CXR images are not the gold standard for pneumonia diagnosis because it has low sensitivity \citep*{hagaman2009admission, self2013high}. Thus, human performance in this problem is usually not very high \citep*{rajpurkar2017chexnet}. That makes us wonder how realistic are some approaches presented in the literature, in which they achieve a very high classification accuracy.

Table \ref{table:classification-parameter} reports the parameters used in the CNN training. We also used a Keras callback to reduce the learning rate by half once learning stagnates for three consecutive epochs.

{
\begin{table}[htbp!]
\centering
\small
\caption{CNN parameters.}
\begin{tabular}{@{}ll@{}}
\toprule
\textit{Parameter} & \textit{Value} \\ \midrule
Warm-up epochs & 50 \\
Fine-tuning epochs & 100 \\
Batch size & 40 \\
Warm-up learning rate & 0.001 \\
Fine-tuning learning rate & 0.0001 \\ \bottomrule
\end{tabular}
\label{table:classification-parameter}
\end{table}
}

\subsubsection{COVID-19 Database (RYDLS-20-v2)}
\label{subsec:dataset}

Table \ref{table:rydls} presents some details of the proposed database, which was named RYDLS-20-v2. The database comprises 2,678 CXR images, with an 80/20 percentage train/test split following a holdout validation split.

Therefore, we performed the split considering some crucial aspects: i) multiple CXR images from the same patient are always kept in the same fold, ii) images from the same source are evenly distributed in the train and test split, and iii) each class is balanced as much as possible while complying with the two previous restrictions. We also created a third set for training evaluation, called validation set, containing 20 percent of the training data randomly.

In this context, given the considerations mentioned above, simple random cross-validation would not suffice since it might not correctly separate the train and test split to avoid data leakage, and it could reduce robustness instead of increasing it. In this context, the holdout validation is a more comfortable option to ensure a fair and proper separation of train and test data. The test set was created to represent an independent test set in which we can validate our classification performance and evaluate the segmentation impact in a less biased context.

{
\begin{table}[htbp!]
\centering
\small
\caption{RYDLS-20-v2 main characteristics.}
\begin{tabular}{@{}llll@{}}
\toprule
\textit{Class} & \textit{Train} & \textit{Validation} & \textit{Test} \\ \midrule
Lung opacity (other than COVID-19) & 739 & 189 & 231 \\
COVID-19 & 315 & 93 & 95 \\
Normal & 673 & 150 & 193 \\ \midrule
\textbf{Total} & \textbf{1727} & \textbf{432} & \textbf{519} \\ \bottomrule
\end{tabular}
\label{table:rydls}
\end{table}
}

We built our database by further expanding our previous work RYDLS-20 \citep*{pereira2020covid} and adopting some guidelines and images provided by the COVIDx dataset \citep*{wang2020covid}. Moreover, we set up the problem with three classes: lung opacity (pneumonia other than COVID-19), COVID-19, and normal. We also experimented with expanding the number of classes to represent a more specific pathogen, such as bacteria, fungi, viruses, COVID-19, and normal. However, in all cases, the trained models did not differentiate between bacteria, fungi, and viruses very well, possibly due to the reduced sample size. Thus, we decided to take a more general approach to create a more reliable classification schema while retaining the focus on developing a more realistic approach.

The CXR images were obtained from eight different sources. Table \ref{table:rydls-distribution} presents the samples distribution for each source.

{
\begin{table}[htbp!]
\centering
\small
\caption{Sources used in RYDLS-20-v2 database.}
\begin{tabular}{@{}p{5.5cm}|lll@{}}
\toprule
\textit{Source} & \textit{Lung opacity} & \textit{COVID-19} & \textit{Normal} \\ \midrule
Dr. Joseph Cohen GitHub Repository \citep*{cohen2020covid} & 140 & 418 & 16 \\
Kaggle RSNA Pneumonia Detection Challenge\footnote{https://www.kaggle.com/c/rsna-pneumonia-detection-challenge} & 1000 & - & 1000 \\
Actualmed COVID-19 Chest X-ray Dataset Initiative\footnote{https://github.com/agchung/Actualmed-COVID-chestxray-dataset} & - & 51 & - \\
Figure 1 COVID-19 Chest X-ray Dataset Initiative\footnote{https://github.com/agchung/Figure1-COVID-chestxray-dataset} & - & 34 & - \\
Radiopedia encyclopedia\footnote{https://radiopaedia.org/articles/pneumonia} & 7 & - & - \\
Euroad\footnote{https://www.eurorad.org/} & 1 & - & - \\
Hamimi's Dataset\citep*{hamimi2016mers} & 7 & - & - \\
Bontrager and Lampignano's Dataset \citep*{ajlan2009swine} & 4 & - & - \\ \bottomrule
\end{tabular}
\label{table:rydls-distribution}
\end{table}
}

We considered posteroanterior (PA) and anteroposterior (AP) projections with the patient erect, sitting, or supine on the bed. We disregarded CXR with a lateral view because they are usually used only to complement a PA or AP view \citep*{bontrager2013textbook}. Additionally, we also considered CXR taken from portable machines, which usually happens when the patient cannot move (e.g., ICU admitted patients). This is an essential detail since there are differences between regular X-ray machines and portable X-ray machines regarding the image quality; we found most portable CXR images in the classes COVID-19 and lung opacity. We removed images with low resolution and overall low quality to avoid any issues when resizing the images.

Finally, we have no further details about the X-ray machines, protocols, hospitals, or operators, and these details impact the resulting CXR image. All CXR images are de-identified\footnote{Aiming at attending to data privacy policies.}, and for some of them, there is demographic information available, such as age, gender, and comorbidities.

\begin{figure}[htbp!]
\centering
\begin{subfigure}{0.32\textwidth}
  \centering
  \includegraphics[width=.95\linewidth]{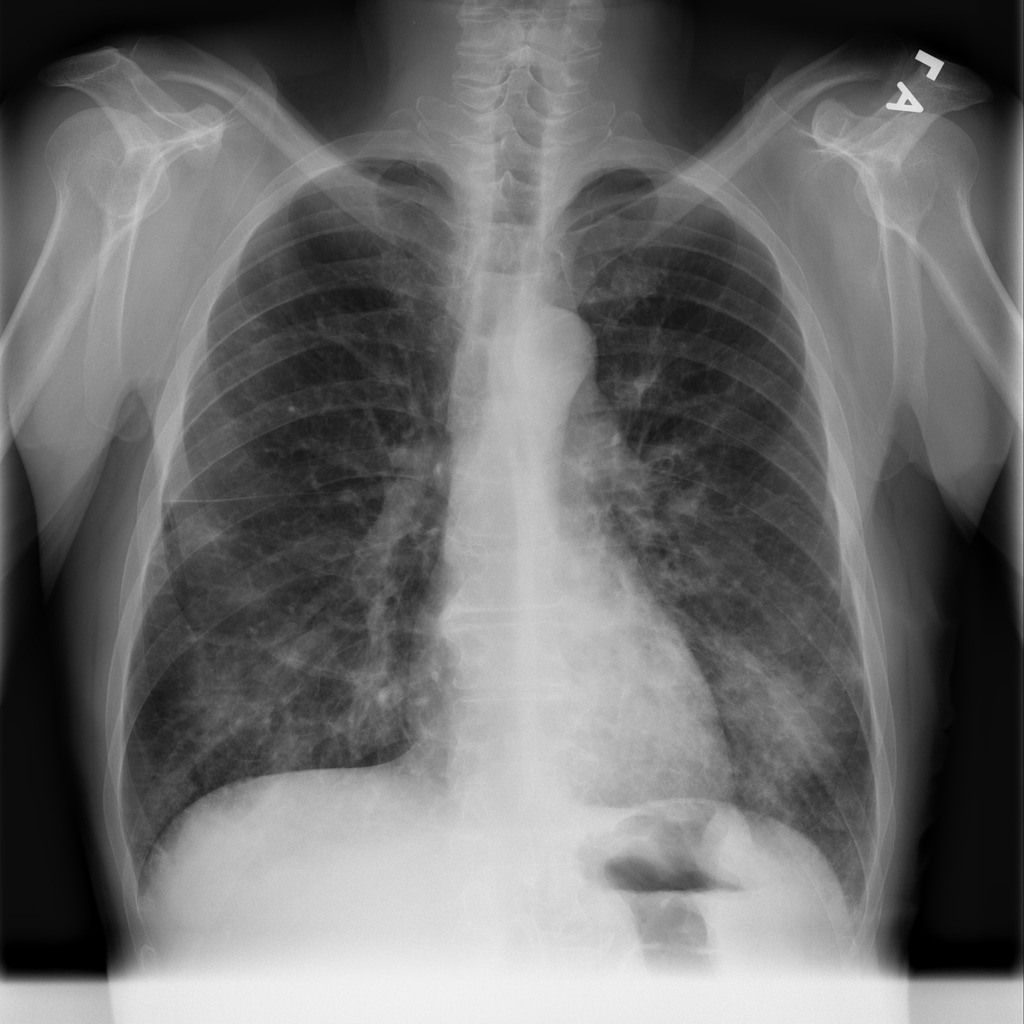}
  \caption{Lung opacity.}
\end{subfigure}
\begin{subfigure}{0.32\textwidth}
  \centering
  \includegraphics[width=.95\linewidth]{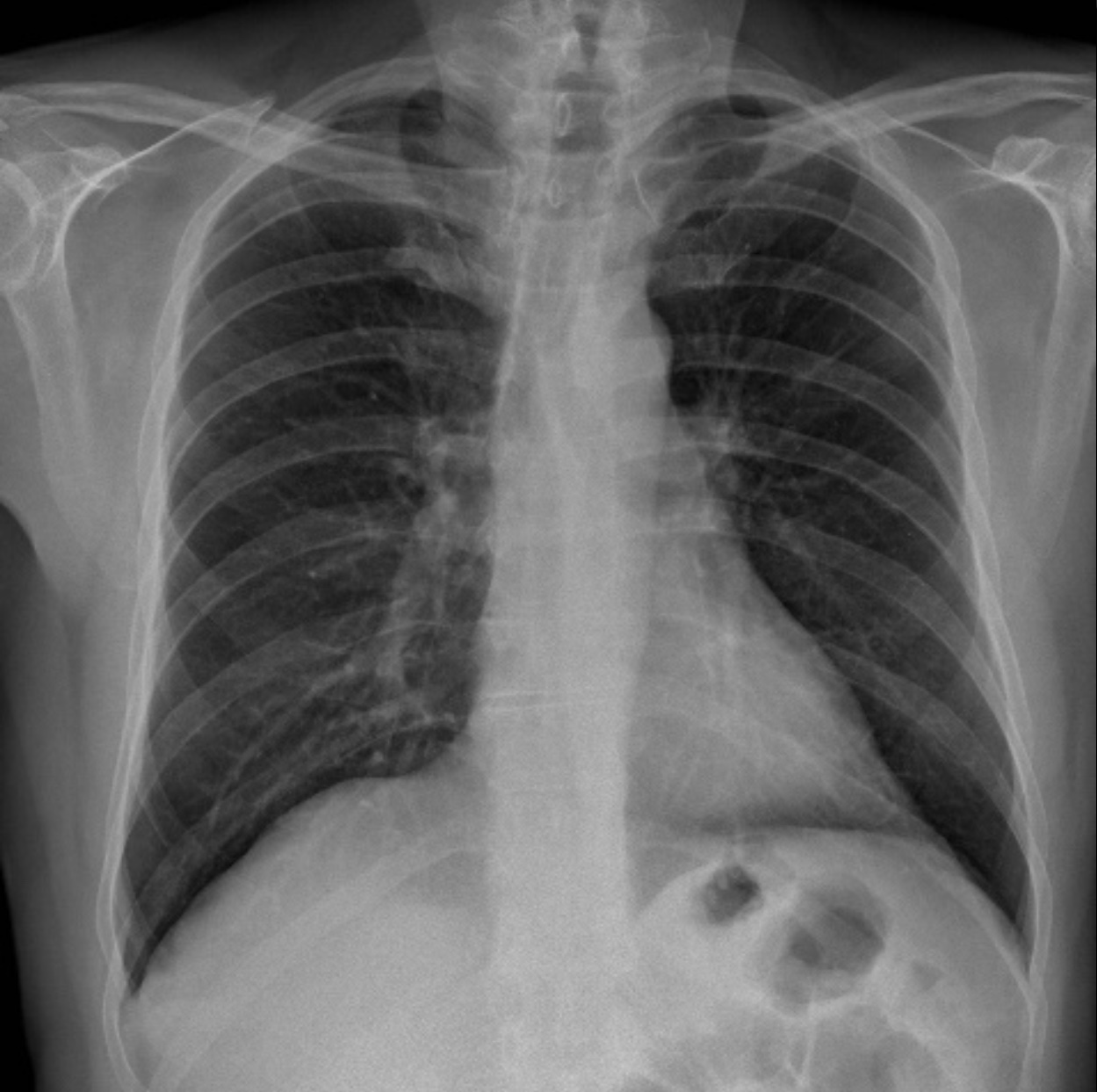}
  \caption{COVID-19.}
\end{subfigure}
\begin{subfigure}{0.32\textwidth}
  \centering
  \includegraphics[width=.95\linewidth]{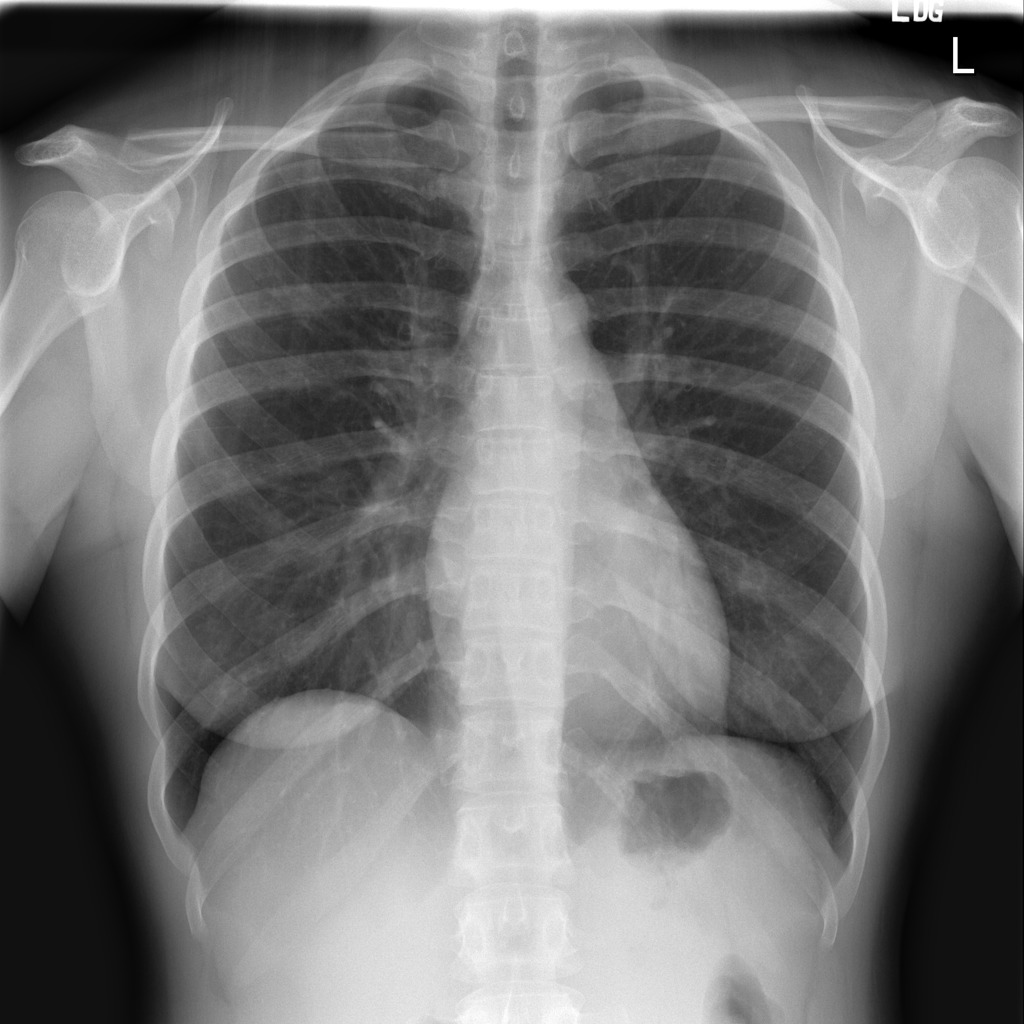}
  \caption{Normal.}
\end{subfigure}
\caption{RYDLS-20-v2 image samples.}
\label{fig:rydls}
\end{figure}

Figure \ref{fig:rydls} presents image examples for each class retrieved from the RYDLS-20-v2 database.

\subsubsection{COVID-19 Generalization}

The COVID-19 generalization intents to demonstrate that our classification schema can identify COVID-19 in different CXR databases. To do so, we set up a binary problem with COVID-19 as the relevant class with a 2-fold validation using only segmented CXR images. The first fold contains all COVID-19 images from the Cohen database and a portion of the RSNA Kaggle database and the second fold contains the remaining RSNA Kaggle database and the other sources. Table \ref{table:covidbias-distribution} shows the samples distribution by source for this experiment. The primary purpose is to evaluate if the CXR images in the Cohen database allows the training of a non-random CNN classifier for the remaining COVID-19 source images and vice versa.

{
\begin{table}[htbp!]
\centering
\small
\caption{COVID-19 generalization database composition.}
\begin{tabular}{@{}p{4cm}llll@{}}
\toprule
\multirow{2}{*}{Source} & \multicolumn{2}{c}{Fold 1} & \multicolumn{2}{c}{Fold 2} \\
\cmidrule{2-3} \cmidrule{4-5}
{} & \textit{Negative} & \textit{COVID-19} & \textit{Negative} & \textit{COVID-19} \\ \midrule
Dr. Joseph Cohen GitHub Repository & 156 & 418 & - & - \\
Kaggle RSNA Pneumonia Detection Challenge & 1000 & - & 1000 & - \\
Actualmed COVID-19 Chest X-ray Dataset Initiative & - & - & - & 51 \\
Figure 1 COVID-19 Chest X-ray Dataset Initiative & - & - & - & 34 \\
Radiopedia encyclopedia & - & - & 7 & - \\
Euroad & - & - & 1 & - \\
Hamimi's Dataset & - & - & 7 & - \\
Bontrager and Lampignano's Dataset & - & - & 4 & - \\
\midrule
\textbf{Total} & 1156 & 418 & 1019 & 85 \\
\bottomrule
\end{tabular}
\label{table:covidbias-distribution}
\end{table}
}

We must highlight that, despite this scenario being our least biased experiment, Kaggle RSNA is used in both folds, so it is not completely bias-free.

\subsubsection{Database Bias}

Moreover, we also evaluated a dataset classification to assess if a CNN can identify the CXR image source using segmented and full CXR images. To do so, we set up a multi-class classification problem with three classes, one for each relevant image source: \texttt{Cohen}, \texttt{RSNA}, and \texttt{Other} (the remaining images from other sources combined). The database comprises 2,678 CXR images, with an 80/20 percentage of train/test split following a random holdout validation split. For training evaluation, we also created a validation set containing 20 percent of the training data randomly. The number of samples distributed among these sets for each data source is presented in Table \ref{table:databasebias-distribution}.

{
\begin{table}[htbp!]
\centering
\small
\caption{Database bias evaluation composition.}
\begin{tabular}{@{}llll@{}}
\toprule
\textit{Class} & \textit{Train} & \textit{Validation} & \textit{Test} \\ \midrule
Cohen          & 364 & 89 & 121 \\
RSNA           & 1288 & 326 & 386 \\
Other          & 61 & 14 & 29 \\ \midrule
\textbf{Total} & \textbf{1713} & \textbf{429} & \textbf{536} \\ \bottomrule
\end{tabular}
\label{table:databasebias-distribution}
\end{table}
}

The rationale is to assess if the database bias is reduced when we use segmented CXR images instead of full CXR images. Such evaluation is of great importance to ensure that the model classifies the relevant class, in this case, COVID-19, and not the image source.

\subsubsection{Data Augmentation}
\label{sec:aug}

\begin{figure}[htbp!]
\centering
\includegraphics[width=.95\linewidth]{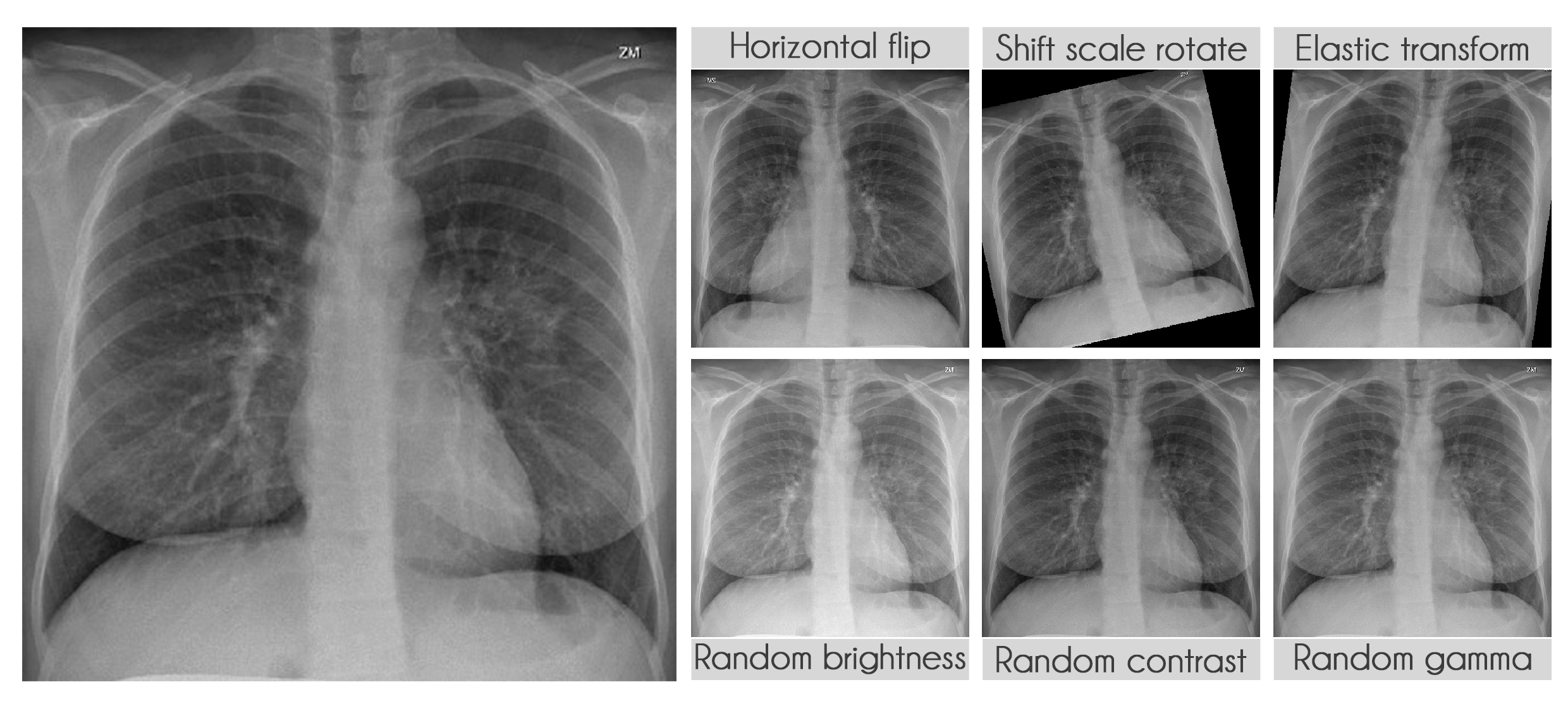}
\caption{Data augmentation examples.}
\label{fig:augmentation-examples}
\end{figure}

We extensively used data augmentation during training in segmentation and classification to virtually increase our training sample size \citep*{shorten2019survey}. Table \ref{table:augmentation-parameter} presents the transformations used during training along with their parameters. The probability of applying each transformation was kept at the default value of 50\%. We used the library \textit{albumentations\footnote{https://github.com/albumentations-team/albumentations}} to perform all transformations \citep*{buslaev2020albumentations}. Figure \ref{fig:augmentation-examples} displays some examples of the transformations applied.

{
\begin{table}[htbp!]
\centering
\small
\caption{Data augmentation parameters.}
\begin{tabular}{@{}lll@{}}
\toprule
\textit{Transformation} & \textit{Segmentation} & \textit{Classification} \\ \toprule
Horizontal flip & -- & -- \\ \midrule
\multirow{3}{*}{Shift scale rotate} & Shift limit = 0.0625 & Shift limit = 0.05 \\
                                    & Scale limit = 0.1    & Scale limit = 0.05 \\
                                    & Rotate limit = 45    & Rotate limit = 15 \\ \midrule
\multirow{3}{*}{Elastic transform} & Alpha = 1         & Alpha = 1 \\
                                   & Sigma = 50        & Sigma = 20 \\
                                   & Alpha affine = 50 & Alpha affine = 20 \\ \midrule
Random brightness & Limit = 0.2 & Limit = 0.2 \\ \midrule
Random contrast & Limit = 0.2 & Limit = 0.2 \\ \midrule
Random gamma & Limit = (80, 120) & Limit = (80, 120) \\ \bottomrule
\end{tabular}
\label{table:augmentation-parameter}
\end{table}
}

\subsection{XAI (Phase 3)}
\label{subsec:lime}

Depending on the perspective, most machine learning models can be seen as a black-box classifier, it receives input and somehow computes an output \citep*{krause2016interacting}. It might happen both with deep and shallow learning, with some exceptions like decision trees. Even though we can measure our model's performance using a set of metrics, it is nearly impossible to make sure that the model focuses on the correct portion of the test image for prediction.

Specifically, in our use case, we want the model to focus exclusively on the lung area and not somewhere else. If the model uses information from other regions, even if very high accuracy is achieved, there can be some limitations to its application, since it is not learning to identify COVID-19 but something else.

\begin{figure}[htbp!]
\centering
\begin{subfigure}{0.4\textwidth}
  \centering
  \includegraphics[width=.98\linewidth]{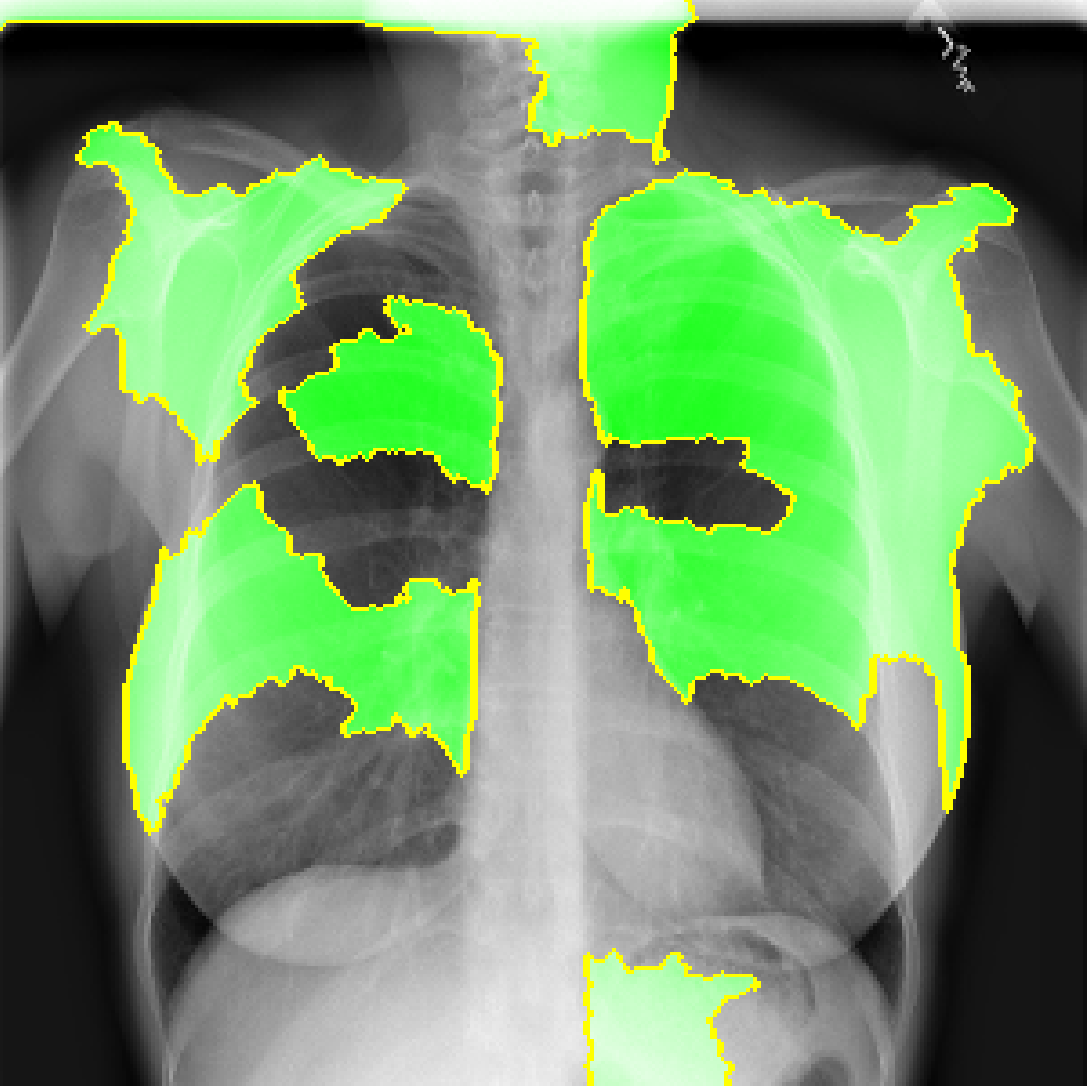}
  \caption{Full CXR image.}
\end{subfigure}
\begin{subfigure}{0.4\textwidth}
  \centering
  \includegraphics[width=.98\linewidth]{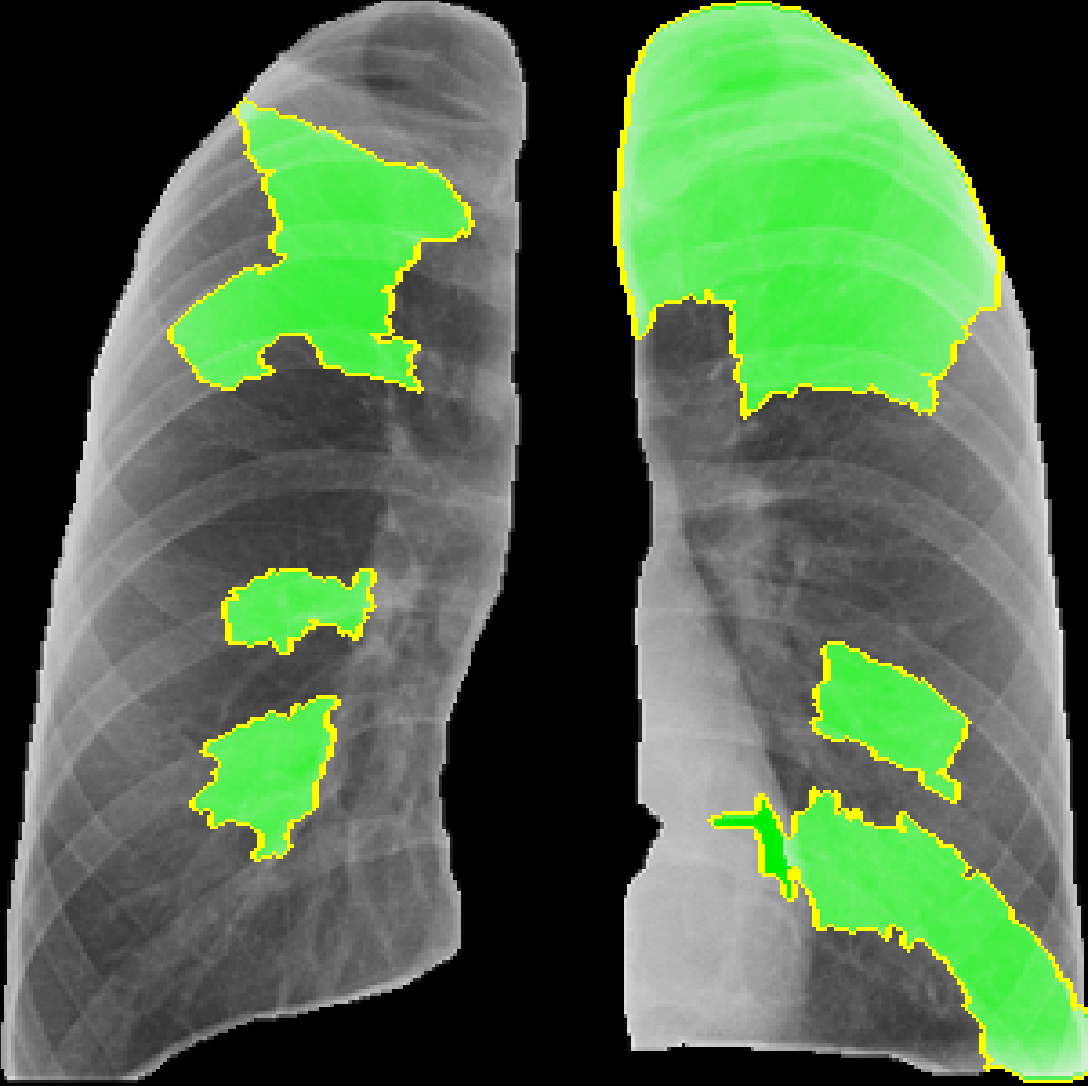}
  \caption{Segmented CXR image.}
\end{subfigure}
\caption{LIME example.}
\label{fig:lime}
\end{figure}

\begin{figure}[htbp!]
\centering
\begin{subfigure}{0.4\textwidth}
  \centering
  \includegraphics[width=.98\linewidth]{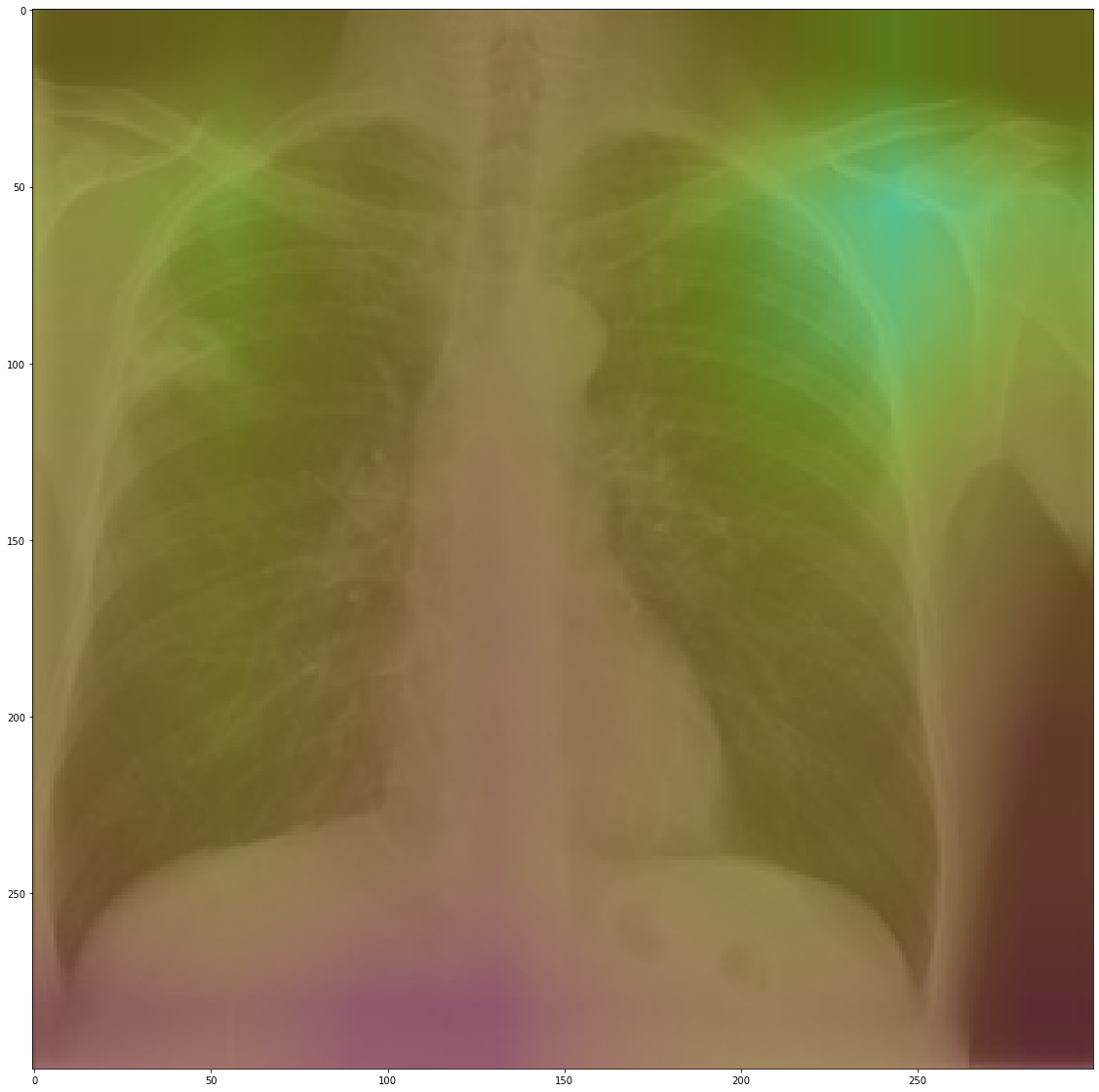}
  \caption{Full CXR image.}
\end{subfigure}
\begin{subfigure}{0.4\textwidth}
  \centering
  \includegraphics[width=.98\linewidth]{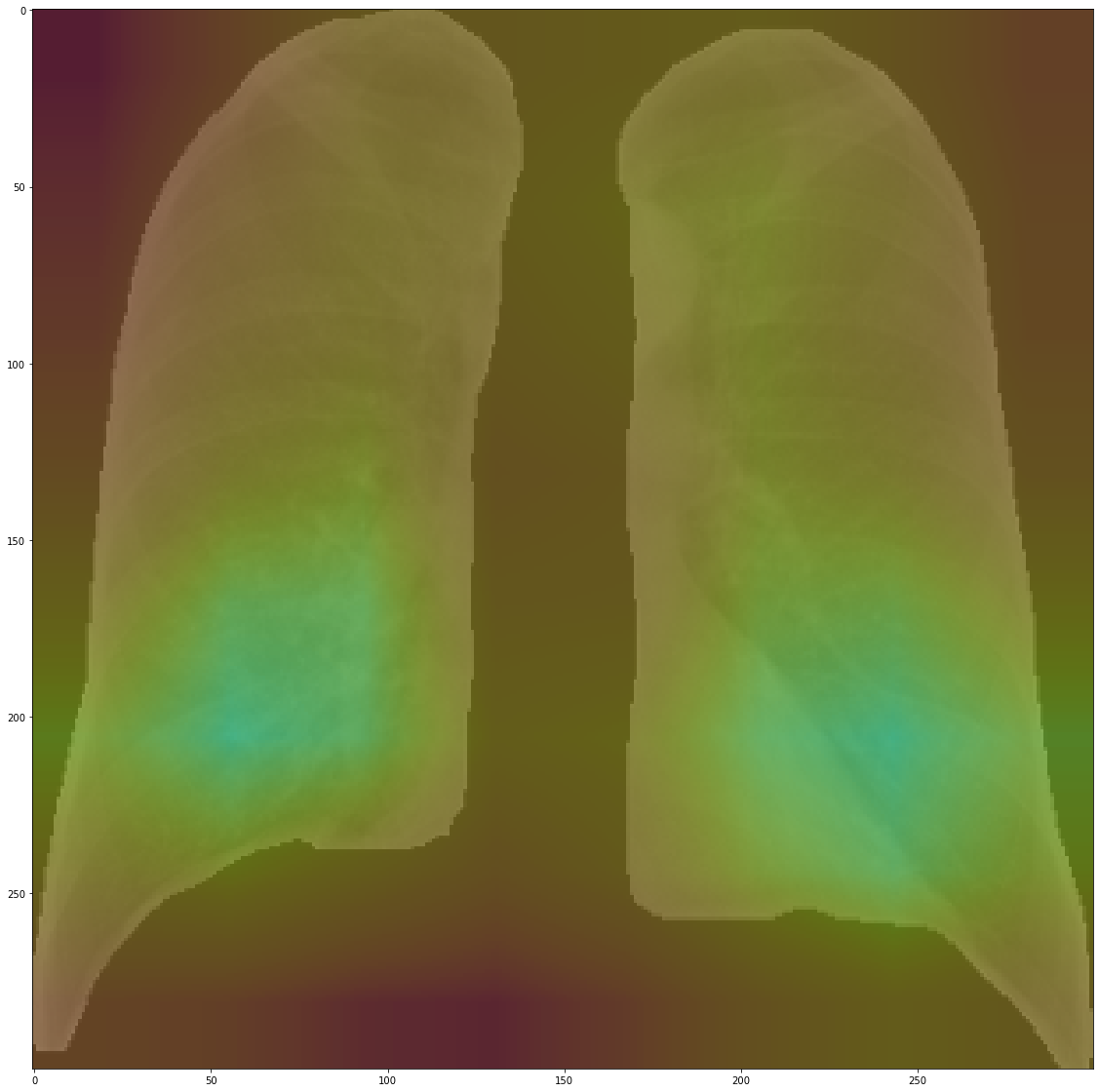}
  \caption{Segmented CXR image.}
\end{subfigure}
\caption{Grad-CAM example.}
\label{fig:gradcam}
\end{figure}

Here, we aim to demonstrate that by using segmented images, the model prediction uses primarily the lung area, which is not often the case when we use full CXR images. To do so, we applied two XAI approaches: LIME and Grad-CAM. Despite having the same main objective, they differ in how they find the important regions. Figures \ref{fig:lime} and \ref{fig:gradcam} shows examples of important regions highlighted by LIME and Grad-CAM, respectively. In section \ref{sec:results}, we will show that models trained using segmented lungs focus primarily on the lung area, while models trained using full CXR images frequently focus elsewhere.

The reason for not using handcrafted feature extraction algorithms here is that it is usually not straightforward to rebuild the reverse path, i.e., from prediction to the raw image. Sometimes, the handcrafted algorithm creates global features, eliminating the possibility of identifying the image regions that resulted in a specific feature.

For each image in the test set, we used LIME and Grad-CAM to find the most important regions used for the predicted class, i.e., regions that support the given prediction. We then summarized all those regions in a heatmap to show the most common regions that the model uses for prediction. Thus, we have one heatmap per classifier per class per XAI approach.

Table \ref{table:lime-parameter} presents the parameters used in LIME. Grad-CAM has a single configurable parameter, which is the convolutional layer to be used, and, in our case, we used the standard approach.

{
\begin{table}[htbp!]
\centering
\small
\caption{LIME parameters.}
\begin{tabular}{@{}ll@{}}
\toprule
\textit{Parameter} & \textit{Value} \\ \midrule
Superpixels identification & Quickshift segmentation \\
Quickshift kernel size & 4 \\
Distance metric & Cosine \\
Number of samples per image & 1000 \\
Number of superpixels in explanation per image & 5 \\
Filter only positive superpixels & True \\ \bottomrule
\end{tabular}
\label{table:lime-parameter}
\end{table}
}

\section{Results}
\label{sec:results}

This section presents an overview of our experimental findings and a preliminary analysis of each contribution individually.

\subsection{Lung Segmentation Results}
\label{sec:results-segmentation}

Table \ref{table:results-segmentation} shows the overall U-Net segmentation performance for the test set for each source we used to compose the lung segmentation database considering the Jaccard distance and the Dice coefficient metrics.

{
\begin{table}[htbp!]
\centering
\small
\caption{Lung segmentation results.}
\begin{tabular}{@{}lll@{}}
\toprule
\textit{Database}      & \textit{Jaccard distance} & \textit{Dice coefficient} \\ \midrule
Cohen v7labs           & 0.041 $\pm$ 0.027         & 0.979 $\pm$ 0.014         \\
Montgomery             & 0.019 $\pm$ 0.007         & 0.991 $\pm$ 0.003         \\
Shenzhen               & 0.017 $\pm$ 0.008         & 0.991 $\pm$ 0.004         \\
JSRT                   & 0.018 $\pm$ 0.011         & 0.991 $\pm$ 0.006         \\
Manually created masks & 0.071 $\pm$ 0.021         & 0.964 $\pm$ 0.011         \\ \midrule
Test set               & 0.035 $\pm$ 0.027         & 0.982 $\pm$ 0.014         \\ \bottomrule
\end{tabular}
\label{table:results-segmentation}
\end{table}
}

As we expected, our manually created masks underperformed when compared to the other sources' results, this may have happened because our masks were not made by professional radiologists. Following that, the Cohen v7labs set also presented a somewhat lower performance. Our manual inspection showed that the model did not include the overlapping region between the lung and heart, and the masks in Cohen v7labs included that region, hence the difference. The performance of the remaining databases is outstanding.

\subsection{Multi-class Classification}
\label{sec:results-classification}

Table \ref{table:results-all} presents F1-Score results for our multi-class scenario. The models using non-segmented CXR images presented better results than the models that used segmented images when we consider raw performance for COVID-19 and lung opacity. Both settings were on par in the normal class.

{
\begin{table}[htbp!]
\centering
\small
\caption{F1-Score results.}
\begin{tabular}{@{}lllll@{}}
\toprule
\textit{Class} & \textit{COVID-19} & \textit{Lung opacity} & \textit{Normal} & \textit{Macro-avg} \\ \midrule
Segmented - VGG16            & 0.83 & 0.88 & 0.9  & 0.87 \\
Segmented - ResNet50V2       & 0.78 & 0.87 & 0.91 & 0.85 \\
Segmented - InceptionV3      & 0.83 & 0.89 & 0.92 & 0.88 \\
\midrule
Non-segmented - VGG16        & 0.94 & 0.91 & 0.91 & 0.92 \\
Non-segmented - ResNet50V2   & 0.91 & 0.9  & 0.92 & 0.91 \\
Non-segmented - InceptionV3  & 0.86 & 0.9  & 0.91 & 0.9  \\
\bottomrule
\end{tabular}
\label{table:results-all}
\end{table}
}

In all cases, the models using segmented images performed worse, considering the selected metric. That result alone might discourage the usage of segmentation in practice. However, in Section \ref{sec:results-lime}, we will show that it is still worth to take into account the segmentation strategy. Even though the use of segmentation does not lead to improvements in the F1-Score rates, the resulting models may present a more realistic performance.

\subsection{COVID-19 Generalization}
\label{sec:results-covidbias}

\begin{figure}[htbp!]
\centering
\includegraphics[width=.5\linewidth]{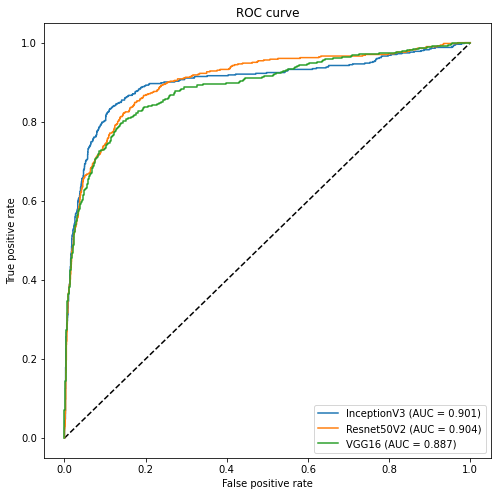}
\caption{COVID-19 Generalization ROC Curve.}
\label{fig:results-covidbias-roc}
\end{figure}

Table \ref{table:results-covidbias} shows the F1-Score results for the COVID-19 generalization. The classification was set up as a binary problem with COVID-19 as the positive class in this problem. The folds were separated in a way that the COVID-19 CXR images from the Cohen database would not be in the same fold of COVID-19 CXR images from the two other databases that contain COVID-19 cases (Actualmed and Figure1 GitHub repositories). The results are auspicious and indeed show that classification, in this case, is far from random. We achieved an F1-Score of 0.77 and 0.7 in the first and second folds, respectively. The lower performance in the second fold was somewhat expected since it contains few COVID-19 examples for training. Figure \ref{fig:results-covidbias-roc} presents the ROC curve for this scenario.

{
\begin{table}[htbp!]
\centering
\small
\caption{F1-Score COVID-19 generalization results.}
\begin{tabular}{@{}llll@{}}
\toprule
\textit{Model}    & \textit{Fold 1} & \textit{Fold 2} & \textit{Macro-avg} \\ \midrule
VGG16             & 0.76            & 0.65            & 0.71 \\
ResNet50V2        & 0.77            & 0.68            & 0.73 \\
InceptionV3       & 0.77            & 0.70            & 0.74 \\
\bottomrule
\end{tabular}
\label{table:results-covidbias}
\end{table}
}

\subsection{Database Bias}
\label{sec:results-databasebias}

Table \ref{table:results-databasebias} shows the F1-Score results for the database bias evaluation. In this problem, the classification was set up as a multi-class problem with database source as the corresponding label for full and segmented CXR images. The results show that overall the lung segmentation reduces the differences between databases. However, even after segmentation, it is possible to identify the source with fair confidence. Such a result may be because the majority of some classes are extracted from the same databases. For instance, most COVID-19 CXR images are from Cohen, and most normal CXR images are from RSNA. Hence in this situation, it is hard to isolate and measure both effects. Furthermore, the class Other contains six different sources, so it is unfair to compare it to Cohen or RSNA. Thus the macro-averaged F1-Score presented does not take it into account. In conclusion, this highlights the need for a bigger and more comprehensive COVID-19 CXR database.

{
\begin{table}[htbp!]
\centering
\small
\caption{F1-Score database bias results.}
\begin{tabular}{@{}lllll@{}}
\toprule
\textit{Scenario} & \textit{Cohen} & \textit{RSNA} & \textit{Other} & \textit{Macro-avg*} \\ \midrule
Segmented - VGG16            & 0.65 & 0.91 & 0    & 0.78 \\
Segmented - ResNet50V2       & 0.62 & 0.9  & 0.07 & 0.76 \\
Segmented - InceptionV3      & 0.61 & 0.89 & 0.24 & 0.75 \\
\midrule
Non-segmented - VGG16        & 0.89 & 0.98 & 0.61 & 0.93 \\
Non-segmented - ResNet50V2   & 0.85 & 0.97 & 0    & 0.91 \\
Non-segmented - InceptionV3  & 0.88 & 0.98 & 0.53 & 0.93 \\ \bottomrule
\multicolumn{5}{l}{*Macro-averaged F1-Score for Cohen and RSNA.}
\end{tabular}
\label{table:results-databasebias}
\end{table}
}

\subsection{XAI Results}
\label{sec:results-lime}

\begin{figure}[htbp!]
\centering
\begin{subfigure}{0.32\textwidth}
  \centering
  \includegraphics[width=.95\linewidth]{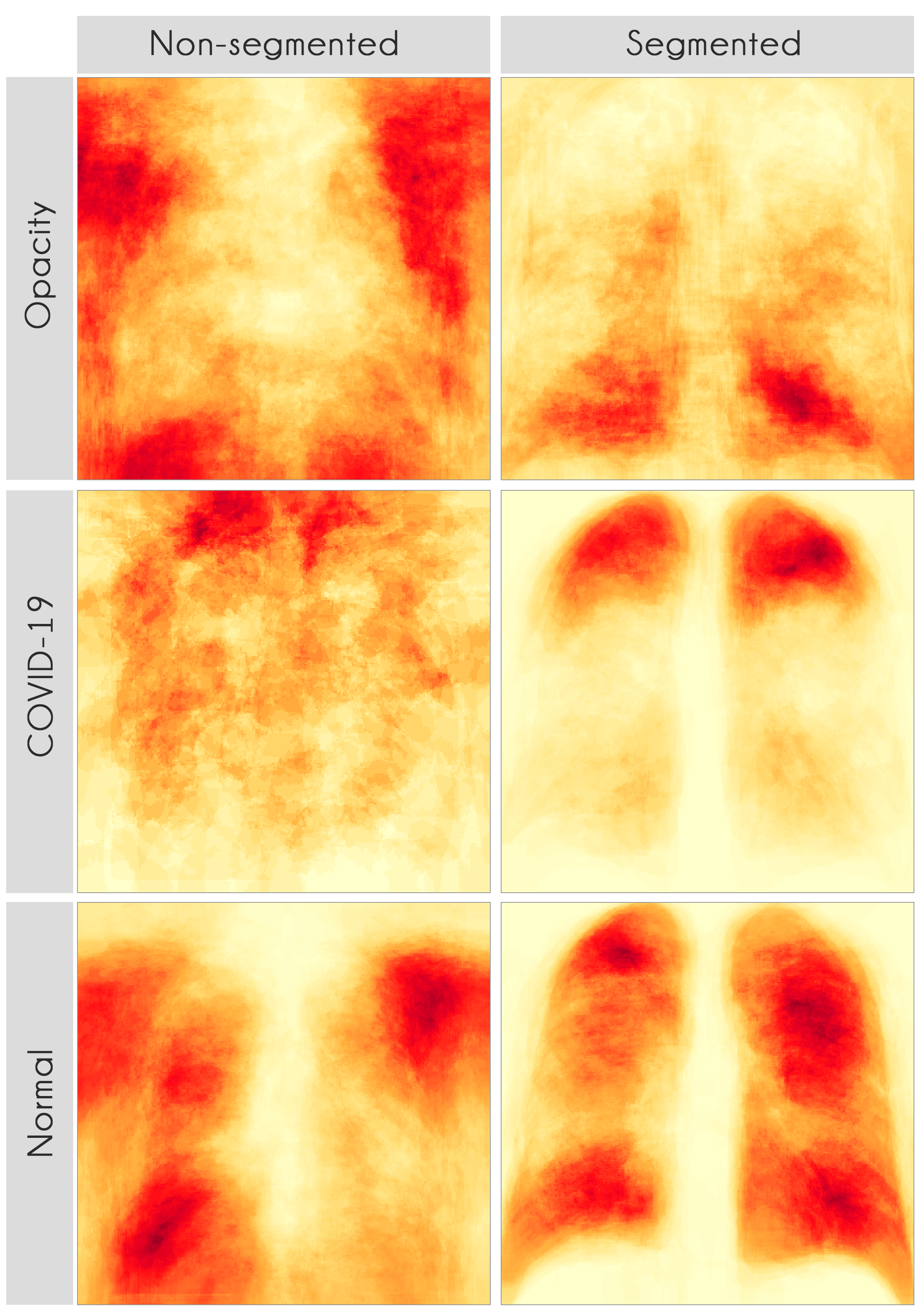}
  \caption{VGG16.}
\end{subfigure}
\begin{subfigure}{0.32\textwidth}
  \centering
  \includegraphics[width=.95\linewidth]{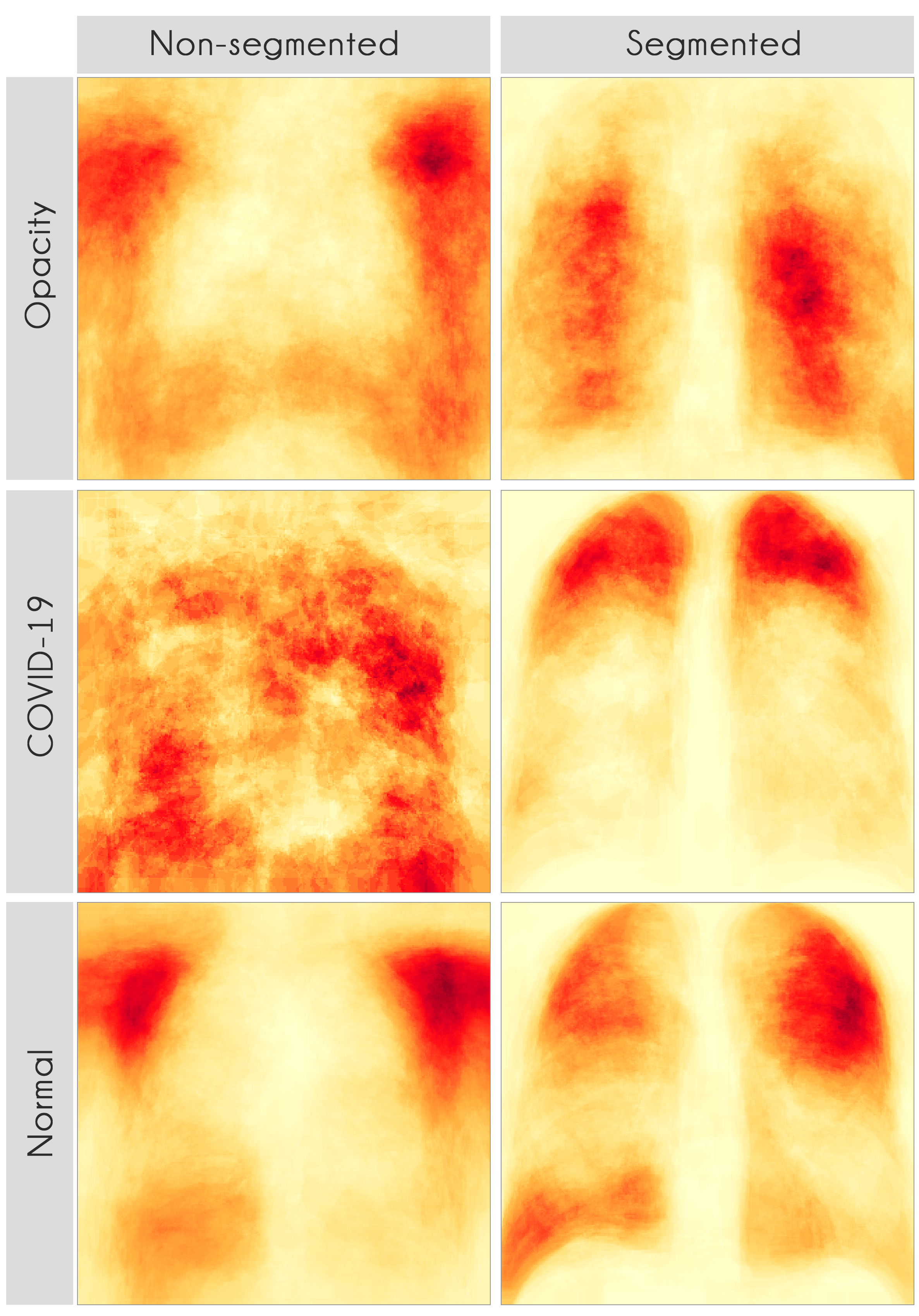}
  \caption{ResNet50V2.}
\end{subfigure}
\begin{subfigure}{0.32\textwidth}
  \centering
  \includegraphics[width=.95\linewidth]{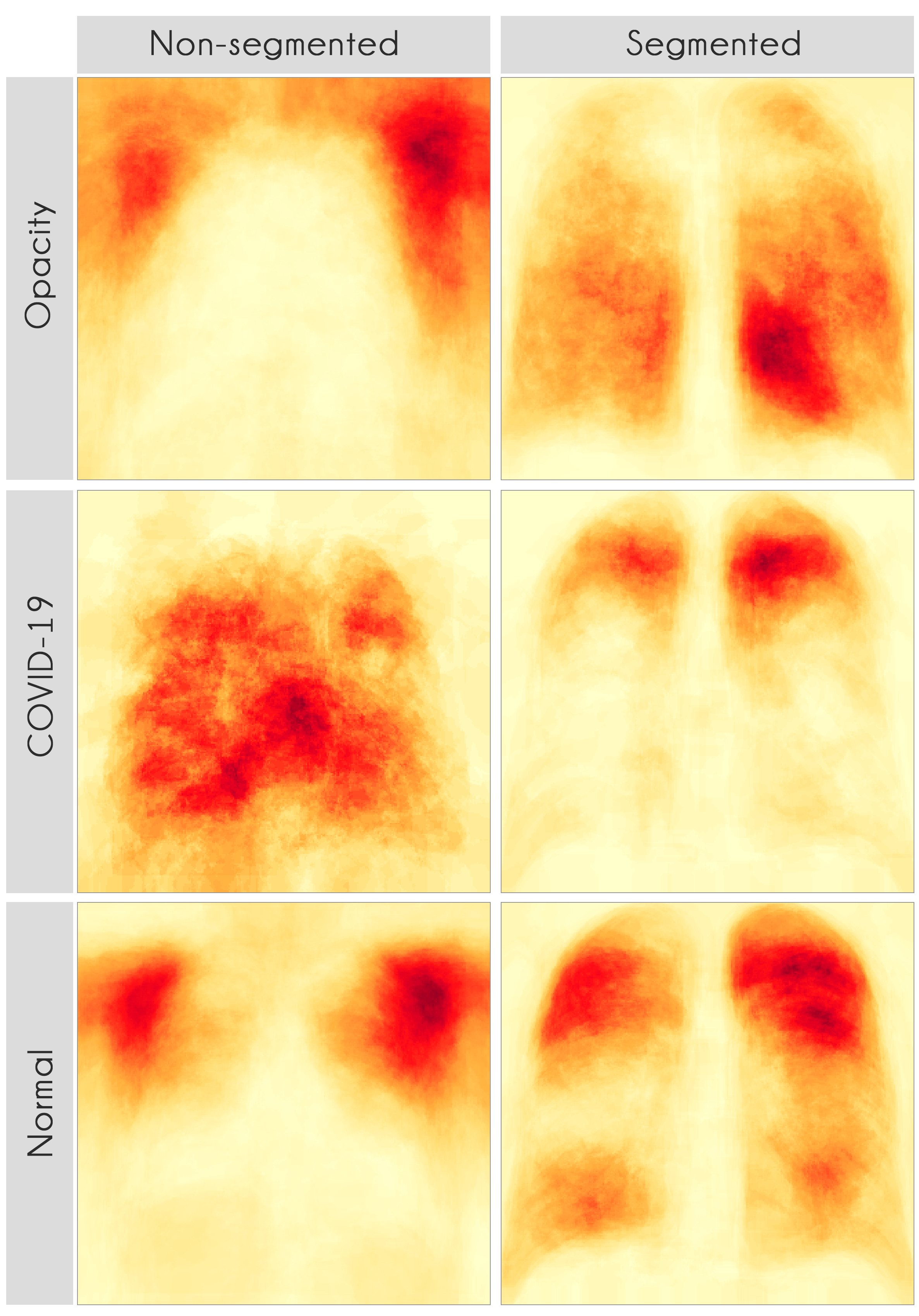}
  \caption{InceptionV3.}
\end{subfigure}
\caption{LIME heatmaps.}
\label{fig:results-lime}
\end{figure}

\begin{figure}[htbp!]
\centering
\begin{subfigure}{0.32\textwidth}
  \centering
  \includegraphics[width=.95\linewidth]{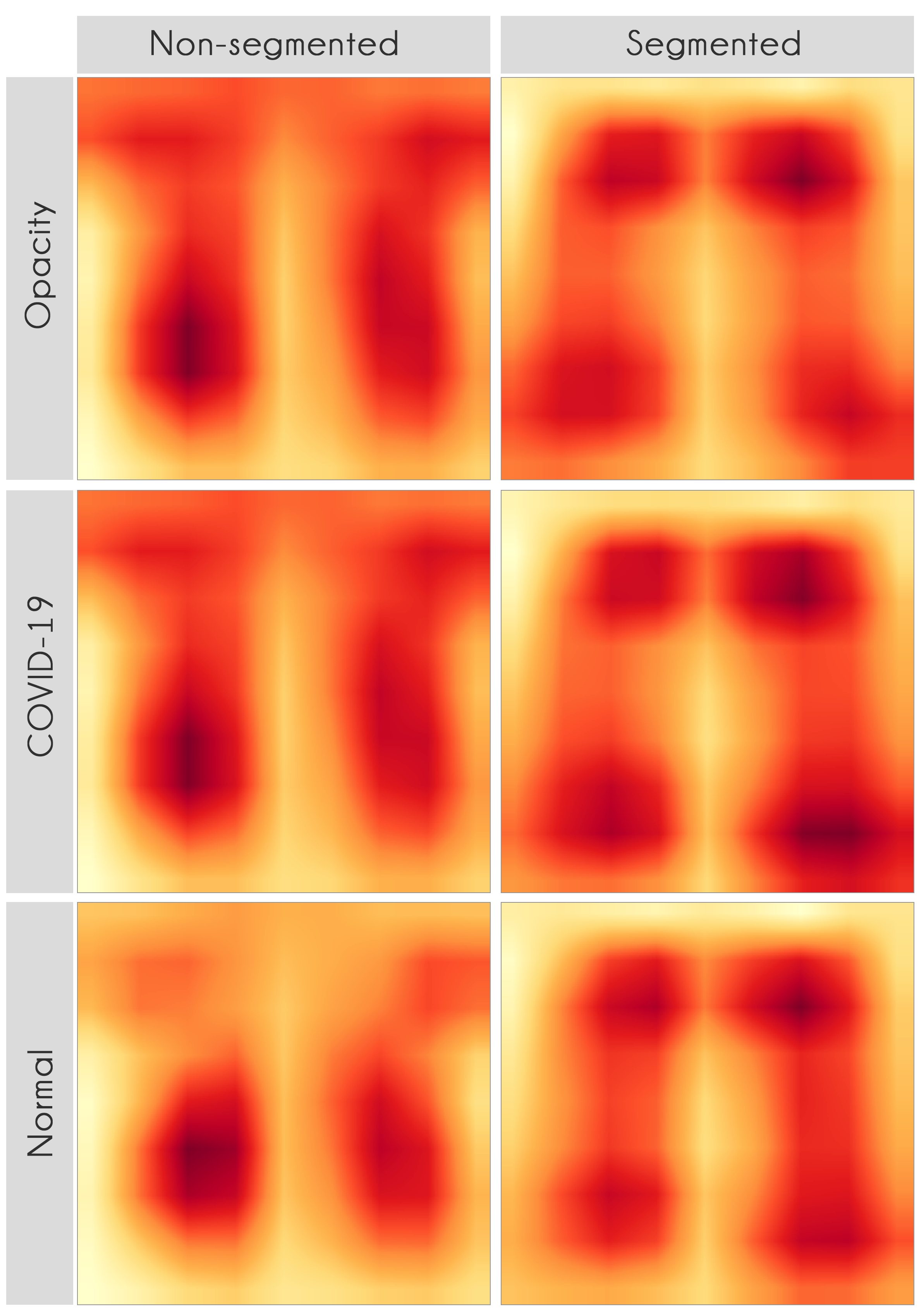}
  \caption{VGG16.}
\end{subfigure}
\begin{subfigure}{0.32\textwidth}
  \centering
  \includegraphics[width=.95\linewidth]{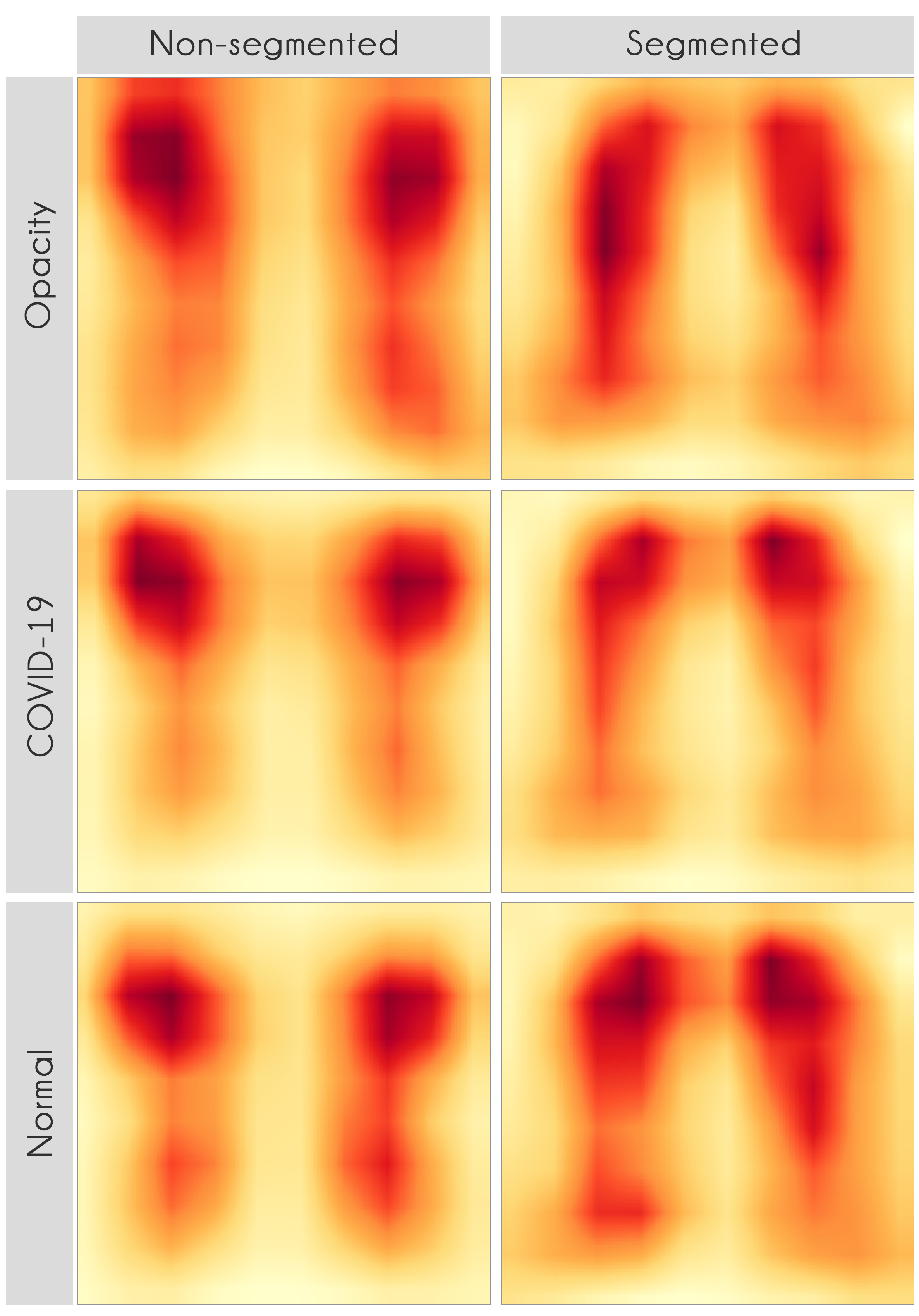}
  \caption{ResNet50V2.}
\end{subfigure}
\begin{subfigure}{0.32\textwidth}
  \centering
  \includegraphics[width=.95\linewidth]{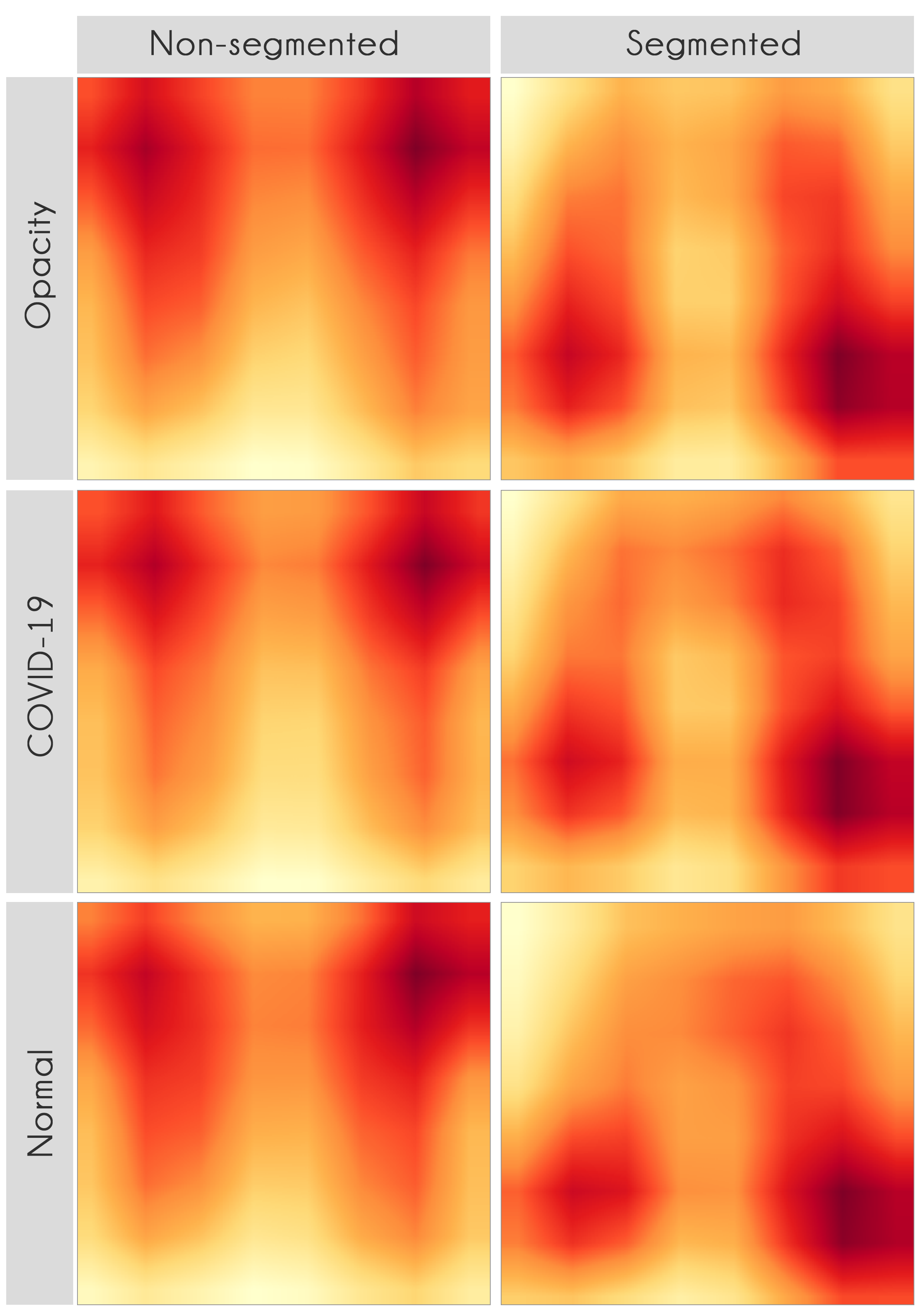}
  \caption{InceptionV3.}
\end{subfigure}
\caption{Grad-CAM heatmaps.}
\label{fig:results-gradcam}
\end{figure}

Figures \ref{fig:results-lime} and \ref{fig:results-gradcam} present the LIME and Grad-CAM heatmaps for our multi-class scenario. We can notice that the models created using segmented CXR images focused primarily in the lung area. The lung shape is discernible in all heatmaps. The only small exception is the VGG16 Lung Opacity class. Despite having the visible lung shape, it also focused a lot in other regions. In contrast, the models that used full CXR images are more chaotic. We can see, for instance, that for both InceptionV3 and VGG16, the Lung Opacity and Normal class heatmaps almost did not focus on the lung area at all.

Even though the models that used full CXR images performed better, considering the F1-Score, they used information outside the lung area to predict the output class. Thus, they did not necessarily learn to identify lung opacity or COVID-19, but something else. Hence, we can say that even though they perform better, considering the classification metric, they are worse and not reliable for real-world applications.

\section{Discussions}
\label{sec:discussion}

This section discusses the importance and significance of the results obtained. Given that we have multiple experiments, we decided to create subsections to drive the discussion better.

\subsection{Multi-class Classification}
\label{sec:discussion-multiclass}

\begin{figure}[htbp!]
\centerline{\includegraphics[width=0.5\columnwidth]{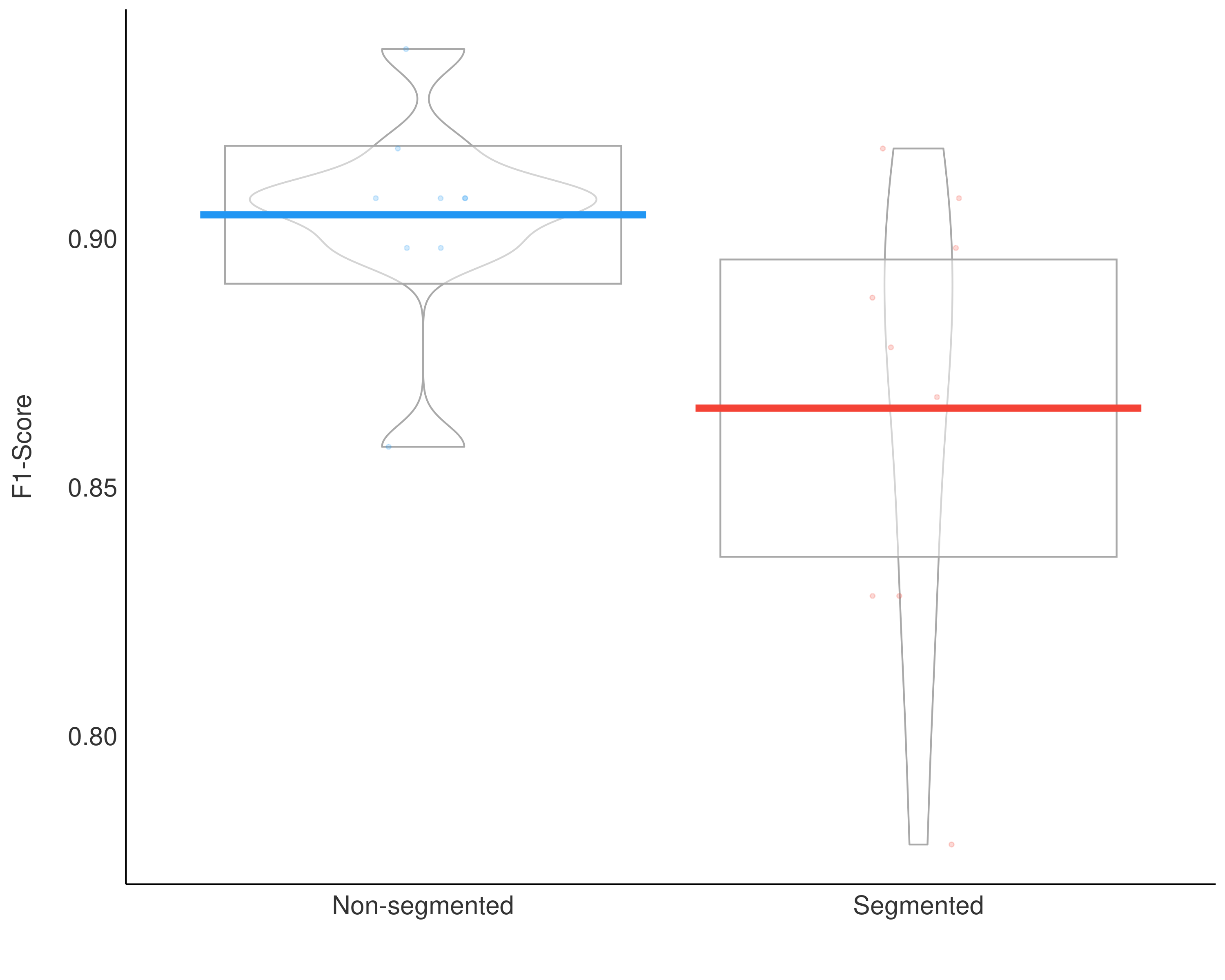}}
\caption{F1-Score results boxplot stratified by segmentation.}
\label{fig:discussion-violin-segm}
\end{figure}

To evaluate the segmentation impact on classification, we applied a Wilcoxon signed-rank test, which indicated that the models using segmented CXR images have a significantly lower F1-Score than the models using non-segmented CXR images ($p=0.019$). Additionally, a Bayesian t-test also indicated that using segmented CXR images reduces the F1-Score with a Bayes Factor of 2.1. The Bayesian framework for hypothesis testing is very robust even for a low sample size \citep{schonbrodt2017sequential}. Figure \ref{fig:discussion-violin-segm} presents a visual representation of our classification results stratified by lung segmentation with a boxplot.

\begin{figure}[htbp!]
\centering
\begin{subfigure}{0.3\textwidth}
  \centering
  \includegraphics[width=.95\linewidth]{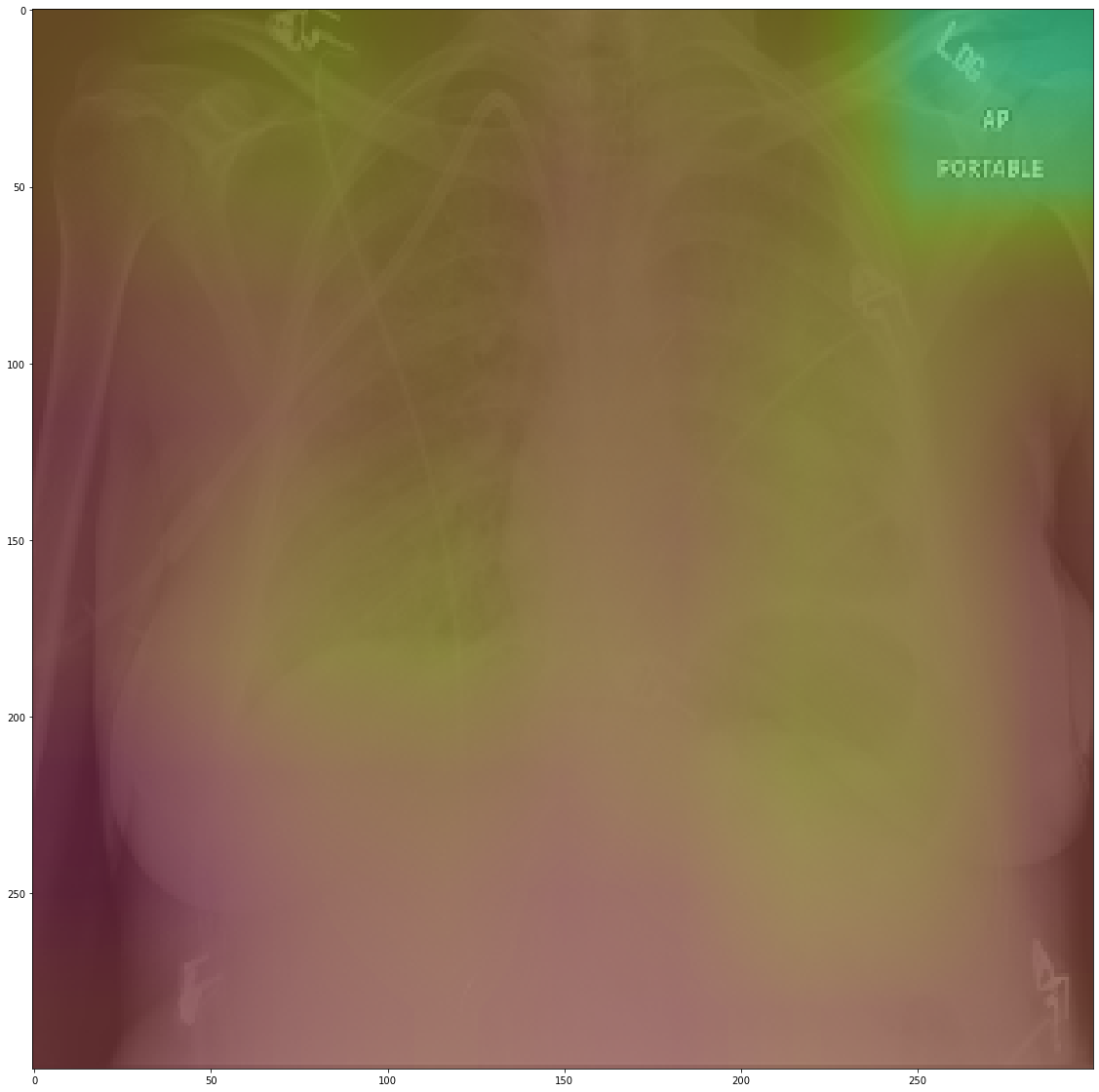}
  \caption{Example 1.}
\end{subfigure}
\begin{subfigure}{0.3\textwidth}
  \centering
  \includegraphics[width=.95\linewidth]{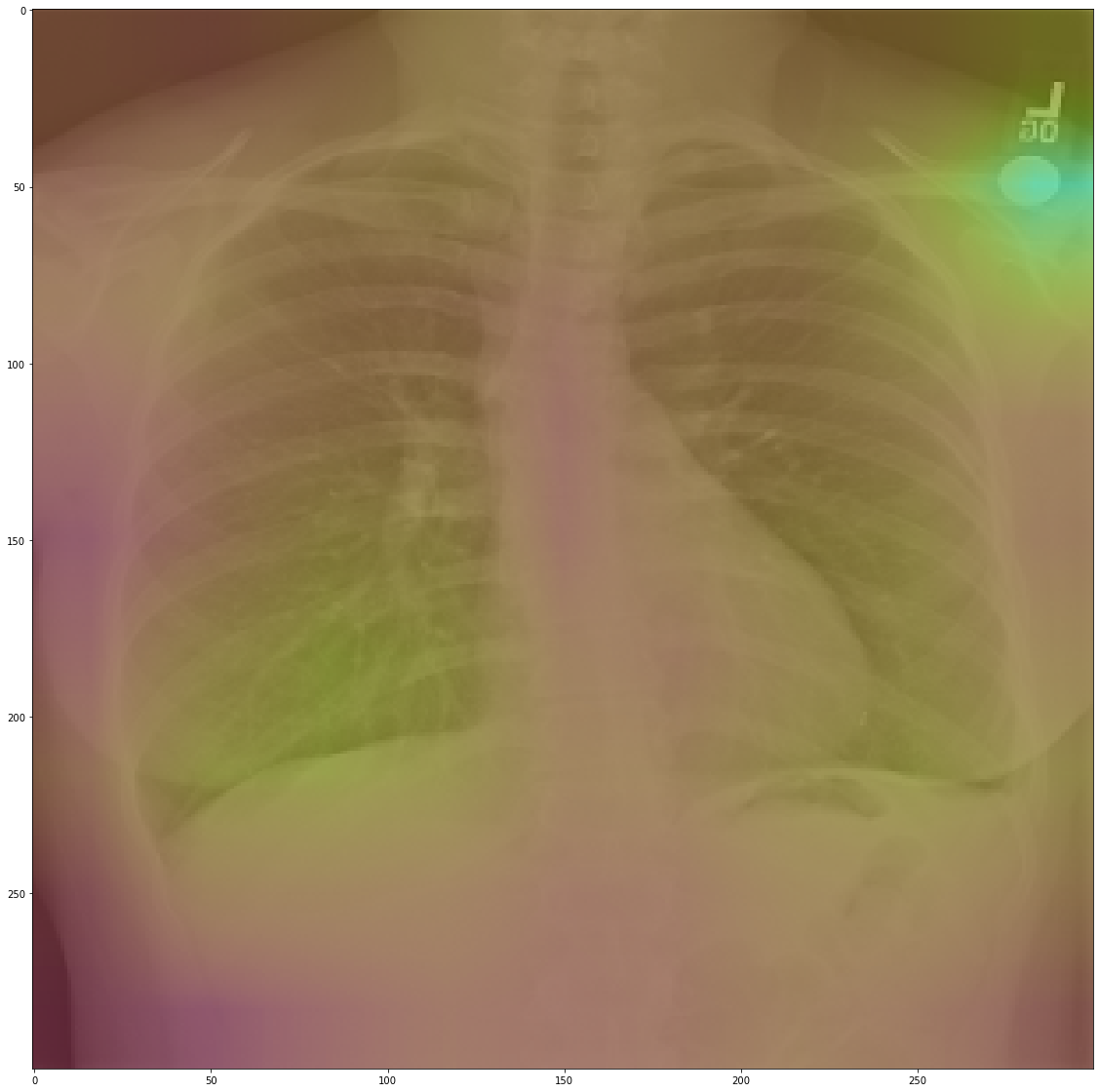}
  \caption{Example 2.}
\end{subfigure}
\caption{Grad-CAM showing a large gradient on CXR annotations.}
\label{fig:discussion-gradcam-bad}
\end{figure}

In general, models using full CXR images performed significantly better, which is an exciting result since we expected otherwise. This result was the main reason we decided to apply XAI techniques to explain individual predictions. Our rationale is that a CXR image contains a lot of noise and background data, which might trick the classification model into focusing on the wrong portions of the image during training. Figure \ref{fig:discussion-gradcam-bad} presents some examples of the Grad-CAM explanation showing that the model is actively using burned in annotations for the prediction. The LIME heatmaps presented in Figure \ref{fig:results-lime} show that exactly behavior for the classes Lung opacity and Normal in the non-segmented models, i.e., the model learned to identify the annotations and not lung opacities. The Grad-CAM heatmaps in Figure \ref{fig:results-gradcam} also show the focus on the annotations for all classes in the non-segmented models.

The most affected class by lung segmentation is the COVID-19, followed by Lung opacity. The Normal class had a minimal impact. The best F1-Scores for COVID-19 and Lung opacity using full CXR images are 0.94 and 0.91, respectively, and after the segmentation, they are 0.83 and 0.89, respectively. We conjecture that such impact comes from the fact that many CXR images are from patients with severe clinical conditions who cannot walk or stand. Thus the medical practitioners must use a portable X-ray machine that produces images with the ``AP Portable'' annotation. That impact also means that the classification models had trouble identifying COVID-19.

\begin{figure}[htbp!]
\centerline{\includegraphics[width=0.5\columnwidth]{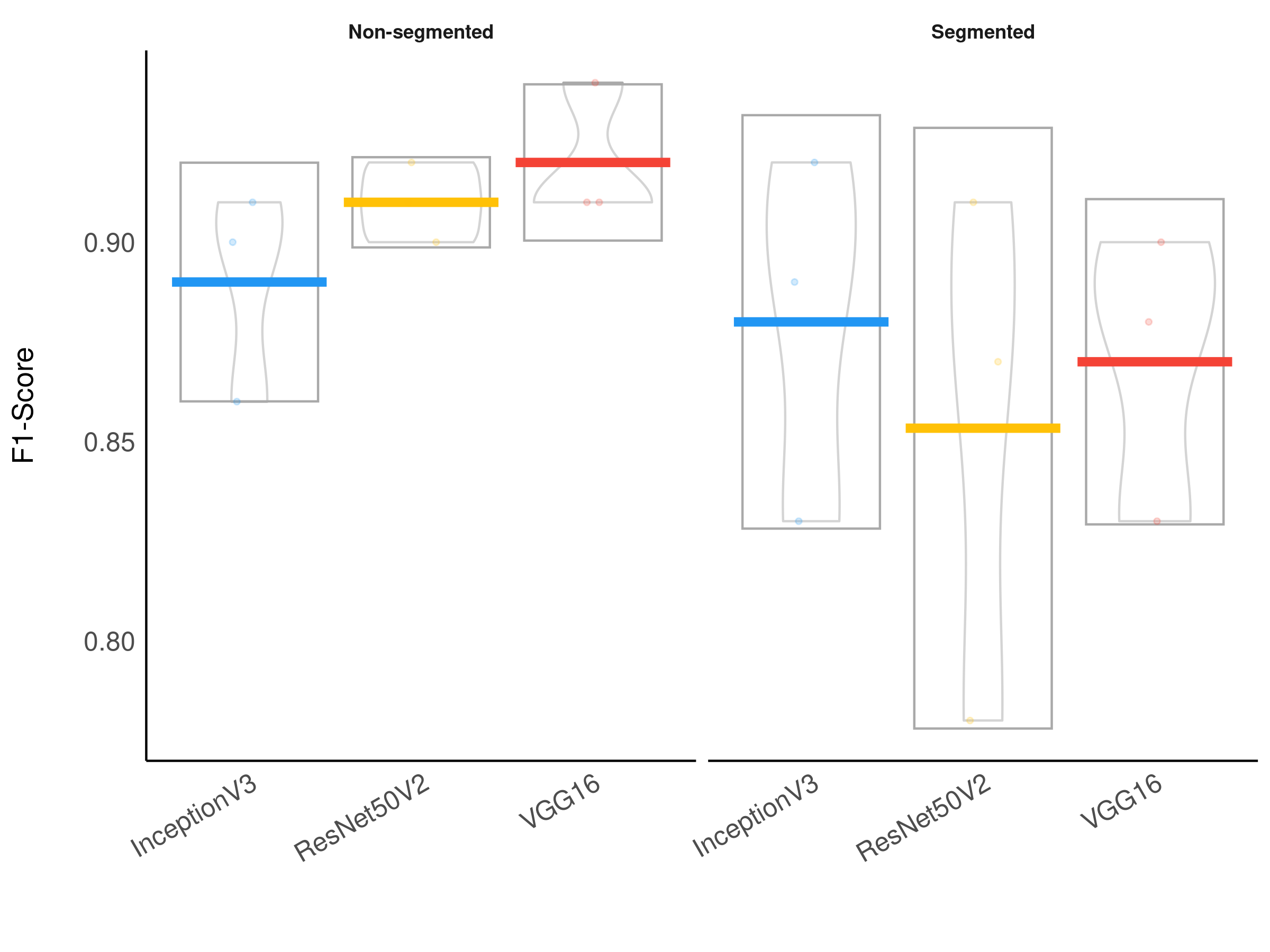}}
\caption{F1-Score results boxplot stratified by segmentation and model.}
\label{fig:discussion-violin-segm-model}
\end{figure}

\begin{figure}[htbp!]
\centerline{\includegraphics[width=0.35\columnwidth]{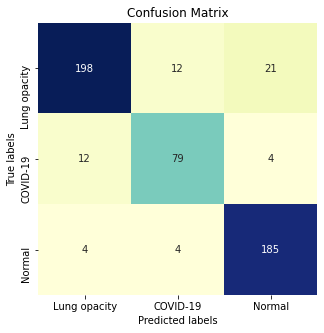}}
\caption{Segmented InceptionV3 Confusion Matrix.}
\label{fig:discussion-conf}
\end{figure}

Considering specifically the models using segmented CXR images, InceptionV3 performed better in all classes. Figure \ref{fig:discussion-violin-segm-model} provides a visual representation of the F1-Score achieved in the experimental results stratified by the model used and lung segmentation. Figure \ref{fig:discussion-conf} shows the confusion matrix for the InceptionV3 using segmented CXR images. Overall the classifier presented a remarkable performance in all labels. The largest misclassification happened with the class Lung opacity being predicted as Normal, followed by the class COVID-19 being predicted as Lung opacity. However, there are reasonable explanations for both: i) Most examples from the classes Lung opacity and Normal came from the RSNA database; thus, we believe that the data source biased the classification marginally; ii) pneumonia caused by COVID-19 could have been confused with pneumonia caused by another pathogen. A solution for both issues would be to increase the number of images in the database, including more data sources.

\subsection{XAI}
\label{sec:discussion-xai}

In this paper, we applied two XAI techniques: LIME and Grad-CAM. The reason for applying both is to evaluate the classification models thoroughly since they work differently. They have some significant differences and highlights: i) LIME is model-agnostic, and Grad-CAM is model-specific; ii) in LIME, the granularity of important regions is correlated to the granularity of the superpixel identification algorithm; iii) Grad-CAM produces a very smoothed output because the dimension of the last convolution layer is much smaller than the dimension of the original input. Keep in mind that such techniques are not definitive. They can complement and corroborate with each other. Thus, we can increase the model reliability in a real-world context by using a more comprehensive approach.

Our XAI approach is novel in the sense that we explored a more general explanation instead of focusing on single examples. In the literature, there are many papers exploring LIME and Grad-CAM for a couple of handpicked examples. The main problem with such approaches is that the examples might have been eventually chosen to reach a specific result. In this paper, we applied the XAI techniques to each image in the test set individually and created a heatmap aggregating all individual results to represent a broader context, which indicates which portions of the CXR image the models have focused on for prediction. Figures \ref{fig:results-lime} and \ref{fig:results-gradcam} demonstrate that the models using full CXR images are misleading because they focus a lot on the left and right uppermost regions, which is usually the location of burned-in annotations.

\subsection{COVID-19 Generalization and Database Bias}
\label{sec:discussion-bias}

The multi-class scenario is fascinating to visualize the behavior of individual models. However, given the strong database bias present in this context, even after lung segmentation, the multi-class results are not entirely reliable.

In order to evaluate such bias and provide a more realistic result, we crafted two specific scenarios to ensure that our classification model is not classifying the database source. First, as we have multiple sources of COVID-19 CXR images, we verified if it was possible to use CXR images from one database to train a model to recognize COVID-19 CXR in the other databases. We achieved a macro-averaged F1-Score of 0.74 using InceptionV3 and an area under the ROC curve of 0.9 using InceptionV3 and ResNet50V2. The F1-Score was lower than in our multi-class scenario. However, this corroborates that it is possible to identify COVID-19 cases across databases, i.e., our classification model is indeed identifying COVID-19 and not the database source. Such a scenario constitutes our main result and contribution, since it represents a less biased and more realistic performance, given the hurdles that still exist with COVID-19 CXR databases.

Second, as discussed in the work of \citep*{maguolo2020critic}, there is a strong bias towards the database source in this context. In our evaluation, we found out that lung segmentation consistently reduces the ability to differentiate the sources. We achieved a database classification F1-Score of 0.93 and 0.78 for full and segmented CXR images, respectively. A Wilcoxon signed-rank test and a Bayesian t-test indicated that segmentation reduces the macro-averaged F1-Score with statistical significance ($p=0.024$ and a Bayes Factor of 4.6). Despite that, even after segmentation, there is a strong bias towards the RSNA Kaggle database, considering specifically this class, we achieved an F1-Score of 0.91. In summary, the usage of lung segmentation is outstanding in reducing the database bias in our context. However, it does remedy the issue entirely.

\subsection{Concluding Remarks}
\label{sec:discussion-conclusion}

In a real-world application, especially in medical practice, we must be cautious and thorough when designing systems aimed at diagnostic support because they directly affect people's lives. A misdiagnosis can have severe consequences for the health and further treatment of a patient. Furthermore, in the COVID-19 pandemic, such consequences can also affect other people since it is a highly infectious disease. Even though the current pandemic attracted much attention from the research community in general, few works focused on a more critical evaluation of the solutions proposed.

Ultimately, we demonstrated that lung segmentation is essential for COVID-19 identification in CXR images through a comprehensive and straightforward application of deep models coupled with XAI techniques. In fact, in our previous work \citep{pereira2020covid}, we have addressed the task of pneumonia identification as a whole, stating that maybe the patterns of the injuries caused by the different pathogens (virus, bacteria, and fungus) are different, so we were able to classify the CXR images with machine learning techniques. Even though the experimental results of that work have shown that it may be possible, it is challenging to be sure that other patterns did not bias the results in the images that were not related to the lungs.

Furthermore, as previously noted, we still believe that even after lung segmentation, the database bias still marginally influenced the classification model. Thus, more aspects regarding the CXR images and the classification model must be further evaluated to design a proper COVID-19 diagnosis system using CXR images.

\section{Conclusion}
\label{sec:conclusion}

The application of pattern recognition techniques has proven to be very useful in many situations in the real world. Several papers propose using machine learning methods to identify pneumonia and COVID-19 in CXR images with encouraging results. However, very few proposed to use lung segmentation to avoid any data leak or overfitting, and only focused on the classification performance.

Considering a real-world application, segmentation is an important step since it removes background information, reduces the chance of data leak, and forces the model to focus only on important image areas. Segmentation might not improve the classification performance, but as it forces the model to use only the lung area information, it increases the model's reliability and quality.

The classification using segmented lungs achieved an F1-Score of 0.88 for the multi-class setup and 0.83 for COVID-19 identification. Using non-segmented CXR images, the classification achieved an F1-Score of 0.92 and 0.94 for the multi-class setup and COVID-19 identification, respectively. The COVID-19 generalization experiment, i.e. using COVID-19 images from one source to predict COVID-19 in a different source, achieved an macro-averaged F1-Score of 0.74.

It is unfair to make direct comparisons of identification rates from different works, as they usually use different databases under different circumstances. Nevertheless, to the best of our knowledge, we achieved the best identification rate of COVID-19 among other types of pneumonia using segmented CXR images. Additionally, we must highlight our novel approach to demonstrate the importance of lung segmentation in CXR image classification.

We do not claim state-of-the-art classification results at this time for a couple of reasons: i) there are some initiatives to build a comprehensive COVID-19 CXR database still ongoing; however, we still do not have a reliable database that can be used as a definitive benchmark; ii) in clinical practice, a small difference in the classification performance is hardly noticeable, and the model reliability and quality are more important than the classification performance \citep*{chen2019develop}; and, iii) the CXR is not the gold standard for diagnosis, even experienced medical practitioners sometimes face doubts when examining a CXR image \citep*{self2013high}; thus we should be very cautious at papers claiming very high classification performance when the human performance is much lower.

Our segmentation approach achieved a Jaccard distance of 0.034 and a Dice coefficient of 0.982, which represents a robust performance considering two factors: i) we did not aim to surpass the state-of-the-art performance of lung segmentation in CXR images; instead, we focused on creating a general segmentation model capable of producing binary lung masks for CXR images in our COVID-19 database; ii) the lung segmentation database was composed of multiple sources, some masks were even manually created. Nevertheless, our approach was on par with current state-of-the-art lung segmentation in CXR images \citep*{chen2018semantic, tang2019xlsor, islam2018towards}.

Furthermore, we applied LIME and Grad-CAM to demonstrate that using segmented CXR images, the models focused primarily on information in the lung area to classify the CXR images. Thus, despite lowering the F1-Score, segmentation improves the prediction quality as it forces the model to use only relevant information.

A potential limitation of this work is the lack of a reliable, definitive COVID-19 CXR database to be used as the benchmark for comparison with the state-of-the-art. Nevertheless, as such, this limitation might also affect the majority of COVID-19 identification works published. Nevertheless, to the best of our knowledge, we achieved the best identification rate of COVID-19 among other types of pneumonia using segmented CXR images in a less biased configuration.

As future work, we aim to keep improving our database to increase our classification performance and provide more robust estimates by using more CNN architectures for segmentation and classification. Furthermore, we want to apply more sophisticated segmentation techniques to isolate specific lung opacities caused by COVID-19. Likewise, we also want to explore more approaches to evaluate the model predictions, such as SHAP \citep*{lundberg2017unified}.

\bibliographystyle{unsrt}
\bibliography{references}

\end{document}